\documentclass[manuscript,screen]{acmart}

\AtBeginDocument{%
  \providecommand\BibTeX{{%
    \normalfont B\kern-0.5em{\scshape i\kern-0.25em b}\kern-0.8em\TeX}}}

\setcopyright{acmcopyright}
\copyrightyear{2018}
\acmYear{2018}
\acmDOI{XXXXXXX.XXXXXXX}

\acmConference[Conference acronym 'XX]{Make sure to enter the correct
  conference title from your rights confirmation emai}{June 03--05,
  2018}{Woodstock, NY}
\acmPrice{15.00}
\acmISBN{978-1-4503-XXXX-X/18/06}

\usepackage{color}
\usepackage{enumitem}
\usepackage{url}
\usepackage{booktabs}
\usepackage{adjustbox}
\usepackage{colortbl}
\usepackage{multirow}
\usepackage{threeparttable}
\usepackage{makecell}
\usepackage{graphicx}

\usepackage{algorithm}
\usepackage{listings}

\usepackage{amsmath}
\usepackage{amsthm}

\newtheorem{definition}{Definition}

\newtheorem{property}{Property}

\usepackage{soul}

\usepackage{pifont}
\newcommand{\cmrk}{\textcolor[rgb]{0,0.7,0}{\ding{51}}}
\newcommand{\xmrk}{\textcolor{red}{\ding{55}}}

\begin{document}

\title{On Elastic Language Models}

\author{Chen Zhang}
\email{czhang@bit.edu.cn}
\affiliation{%
  \institution{Beijing Institute of Technology}
  \country{China}
}

\author{Benyou Wang}
\authornote{Benyou Wang and Dawei Song are corresponding authors.}
\email{larst@affiliation.org}
\affiliation{%
  \institution{The Chinese University of Hong Kong, Shenzhen}
  \country{China}
}

\author{Dawei Song}
\authornotemark[1]
\email{dwsong@bit.edu.cn}
\affiliation{%
  \institution{Beijing Institute of Technology}
  \country{China}
}

\renewcommand{\shortauthors}{Zhang, et al.}

\begin{abstract}
Large-scale pretrained language models have achieved compelling performance in a wide range of language understanding and information retrieval tasks. While the large scales ensure capacity, they also hinder deployment. Knowledge distillation offers an opportunity to compress a large language model to a small one, in order to reach a reasonable latency-performance tradeoff. However, for scenarios where the number of requests (e.g., queries submitted to a search engine) is highly variant, the static tradeoff attained by the compressed language model might not always fit. Once a model is assigned with a static tradeoff, it could be inadequate in that the latency is too high when the number of requests is large or the performance is too low when the number of requests is small. To this end, we propose an elastic language model (\textsc{ElasticLM}) that elastically adjusts the tradeoff according to the request stream. The basic idea is to introduce a compute elasticity to the compressed language model, so that the tradeoff could vary on-the-fly along scalable and controllable compute. Specifically, we impose an elastic structure to enable \textsc{ElasticLM} with compute elasticity and design an elastic optimization to learn \textsc{ElasticLM} under compute elasticity. To serve \textsc{ElasticLM}, we apply an elastic schedule. Considering the specificity of information retrieval, we adapt \textsc{ElasticLM} to dense retrieval and reranking and present \textsc{ElasticDenser} and \textsc{ElasticRanker} respectively. Offline evaluation is conducted on a language understanding benchmark GLUE; and several information retrieval tasks including Natural Question, Trivia QA, and MS MARCO. The results show that \textsc{ElasticLM} along with \textsc{ElasticDenser} and \textsc{ElasticRanker} can perform correctly and competitively compared with an array of static baselines. Furthermore, online simulation with concurrency is also carried out. The results demonstrate that \textsc{ElasticLM} can provide elastic tradeoffs with respect to varying request stream. We will release our code and checkpoints for guaranteed reproducibilty.
\end{abstract}



\received{20 February 2007}
\received[revised]{12 March 2009}
\received[accepted]{5 June 2009}

\maketitle

\section{Introduction}

Pretrained language models (LMs)~\citep{DevlinCLT19} have been applied to a wide range of language understanding~\citep{WangSMHLB19,RaffelSRLNMZLL20} and information retrieval (IR) tasks~\citep{NguyenRSGTMD16,Ni21}, and achieved remarkable performance. However, they are inherently inefficient due to their large scales. Typically, a larger LM would lead to improved performance at the cost of prolonged latency.

\begin{figure}[t]
    \centering
    \includegraphics[width=0.55\textwidth]{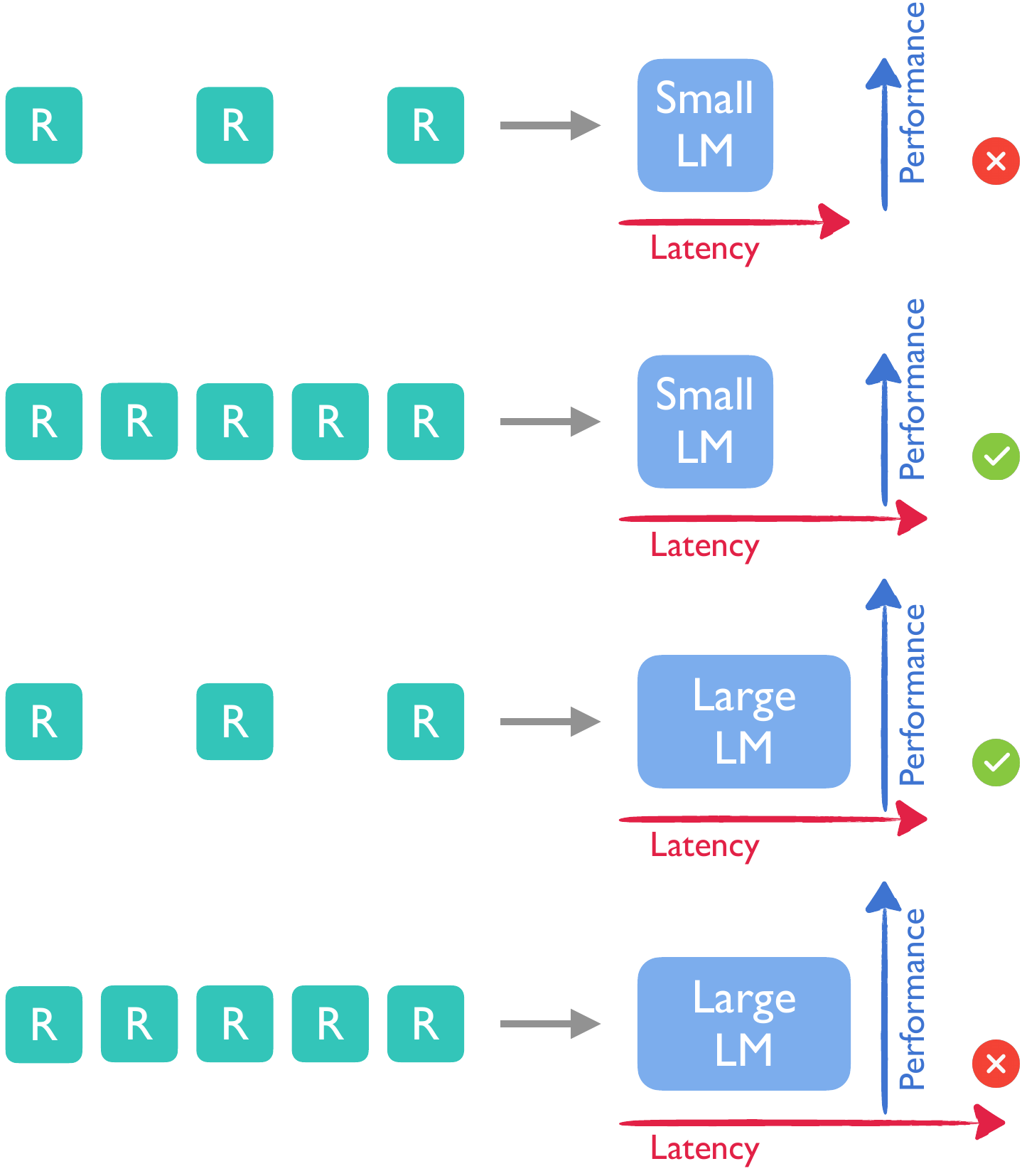}
    \caption{Tradeoffs hidden beneath small and large LMs do not fit well for all cases. A large LM would suffer from high latency when the number of requests is large, and a small LM would suffer from low performance when the number of requests is small.}
    \label{fig:1}
\end{figure}

Knowledge distillation~\citep{HintonVD15}, as an alternative to model quantization~\citep{SungSH15} and sparsification~\citep{HanPTD15}, can alleviate the concern by distilling a large LM into one of a small one and realizing a reasonable tradeoff between reduced latency and acceptable performance. However, such static tradeoff might not fit well for all cases. The reason lies in that the number of requests (e.g., queries submitted to a search system concurrently) would be variant from one time to another. For example, the number of users engaged in a real-world application would sometimes reach a peak but routinely stay very small. Once a model is assigned with a static tradeoff, it could be inadequate in that the latency is too high when the number of requests is large or the performance is too low when the number of requests is small, as shown in Figure~\ref{fig:1}. In this case, an elastic tradeoff is favored to elastically balance the latency and performance, such that the latency should not be high when the number of requests is large while the performance should not be low when the number of requests is small.

To meet the demand, we propose an elastic language model, namely \textsc{ElasticLM}, which is incorporated with knowledge distillation and is subject to tradeoff variation. To realize the elastic tradeoff, \textsc{ElasticLM} requires a compute elasticity, where the tradeoff could vary on-the-fly along scalable and controllable compute. To enable \textsc{ElasticLM} with compute elasticity, we use an elastic structure that can interpret a LM as an ensemble of \textsc{SubLM}s. In the structure, certain parameters can be dropped to realize a small \textsc{SubLM}, but can also be added back to recover a large \textsc{SubLM}. To learn \textsc{ElasticLM} under compute elasticity, we then design an elastic optimization, in which all the defined \textsc{SubLM}s will be traversed and learned in one optimization step. After that, \textsc{ElasticLM} can be driven by an elastic schedule in application of compute elasticity with respect to the number of queued requests. Considering that IR is a scenario with rather tremendous concurrency~\citep{FanXCCMLZG22,LinNY21}, we further adapt \textsc{ElasticLM} to dense retrieval and reranking, separately giving rise to \textsc{ElasticDenser} and \textsc{ElasticRanker}. To our best knowledge, \textsc{ElasticLM} is the first work that explores to own an elastic tradeoff, while \textsc{ElasticDenser} is further the first work that aims to compress dense retrievers.

Offline evaluation is conducted on a language understanding benchmark (GLUE) for \textsc{ElasticLM}; two open-domain question answering datasets (Natural Question and Trivia QA) and a passage ranking dataset (MS MARCO Passage) for \textsc{ElasticDenser}; and a document ranking dataset (MS MARCO Document) for \textsc{ElasticRanker}. The results show that \textsc{ElasticLM} along with \textsc{ElasticDenser} and \textsc{ElasticRanker} can perform correctly with compute elasticity and are competitive with static baselines. We also carry out online simulation with concurrency. The results demonstrate that \textsc{ElasticLM} can enable elastic tradeoffs in comparison with static baselines.

To sum up, our contributions can be summarized as follows:
\begin{itemize}
    \item We identify that a static latency-performance tradeoff would not always fit for all cases. This is to our best knowledge the first time that the problem is recognized and formulated.
    \item To provide an elastic tradeoff, we propose \textsc{ElasticLM} with compute elasticity as the most basic intuition. The proposal is shown to be the first trial in related area.
    \item \textsc{ElasticLM} is further specialized as \textsc{ElasticDenser} and \textsc{ElasticRanker} for IR, suggesting the possibility of adapting \textsc{ElasticLM} to a few more scenarios.
    \item The evaluation and simulation results showcase that \textsc{ElasticLM} is empirically correct and competitive, and exactly owns an elastic tradeoff.
\end{itemize}

\section{Related Work}

\subsection{Knowledge Distillation}

Prior studies have explored ways to make LMs small under the regime of teacher-student distillation~\citep{HintonVD15}, for both language understanding~\citep{SunCGL19,HouHSJCL20,JiaoYSJCL0L20} and IR~\citep{GaoDC20,ChenHHSS21,XinNYL20}. It is noteworthy that knowledge distillation for model compression in IR is studied only for reranking but not for first-stage retrieval. However, knowledge distillation for model generalization in IR spans a wide range, which appear mainly in the literature of dense retrieval.

Generally, LM distillation contains two lines of work: task-specific distillation and task-agnostic distillation. The former~\citep{SunCGL19,HouHSJCL20,Yang22,Zhang22} conducts distillation from a finetuned LM over the data of a specific task. The latter~\citep{JiaoYSJCL0L20,WangW0B0020,WangBHDW21} conducts distillation from a pretrained LM over the data used for pretraining. It is commonly recognized that task-agnostic distillation leads to better performance due to its better feature alignment ability. Besides the data in use, the existing methods mainly differ in their distillation objectives. For example, PKD~\citep{SunCGL19} urges the need of hidden states, and TinyBERT~\citep{JiaoYSJCL0L20} highlights the importance of attention scores.

From another perspective, LM distillation methods can also be categorized into two types: once-for-one and once-for-all. The once-for-one methods initialize the student in a static style. Typically, the initialization can be realized by borrowing structures from the teacher~\citep{SunCGL19} or by complete randomness~\citep{WangW0B0020}. The limitation of the once-for-one methods is recognized that they need to be learned for different devices with different compute constraints, in contrast, the once-for-all methods~\citep{CaiGWZH20,HouHSJCL20,FanGJ20,KimC20,WangWLCZGH20,LiuSHWWZJCHQ22} gain intuition from pruning~\citep{HanPTD15,MichelLN19,YangZWS22,MaZ22}. The basic idea is that a small model is a subset of a large model in the context of pruning, since the small model can be obtained by pruning the large model. In return, the large model is a superset of the small model and should share its parameters to the small model. Thereby, the student is pruned to an ensemble of multiple submodels in a parameter-sharing fashion. For example, DynaBERT~\citep{HouHSJCL20} shares attention heads and feedforward neurons, LayerDrop~\citep{FanGJ20} shares layers, and LengthDrop~\citep{KimC20} shares tokens. In this circumstance, these submodels can be learned in one run and be used flexibly according to concerned devices.

Although the once-for-all distillation in fact achieves offline scalable compute, little work has paid attention to the online scalable compute. Early Exiting can be viewed as the only existing approach to achieve online scalable compute~\citep{XinTLYL20,ZhouXGM0W20,LiuZWZDJ20,LiuSHWWZJCHQ22}. It controls the exit of the computation adaptively based on the input. Unfortunately, it lacks a mechanism for controllable compute conditional on varying request steam. Our work is motivated by the once-for-all methods, and aims to derive an elastic structure that contain an ensemble of \textsc{SubLM}s in one LM, and an elastic optimization to pretrain the LM through distillation. With a dedicated elastic schedule, our elastic LM can serve online with scalable and controllable compute with respect to varying request stream.

Note that an elastic compute could be an alternative to what we want to accomplish in an elastic LM. However, the elastic compute can introduce more compute (e.g., more devices) according to the number of requests, thus would introduce additional memory footprint. On the contrary, the proposed elastic LM would only sacrifice performance at an acceptable level for latency.

\subsection{Dense Retrieval}

Compared with conventional IR models (e.g., BM25~\citep{RobertsonZ09}) that only consider lexical interactions between a query and a document, neural IR models (e.g., Duet~\citep{Mitra0C17}) incorporate the semantic interaction, yielding increased performance yet higher latency. There are two branches of methods in neural IR, i.e., sparse retrieval that strives to enhance sparse term representations, and dense retrieval that aims to improve dense text representations. 

The sparse retrieval methods yield either a better term reweighting~\citep{DaiC20,GaoDC21} or better document expansion~\citep{Nogueira19a,Nogueira19b,FormalPC21}. For example, DeepCT~\citep{DaiC20} leverages contextualized representations to compute term weights, and COIL~\citep{GaoDC21} enables the inverted list with contextual information. Doc2Query~\citep{Nogueira19a} expands a document by predicting associated queries. DocTTTTTQuery~\citep{Nogueira19b} is an improved version of Doc2Query by employing a pretrained LM T5~\citep{RaffelSRLNMZLL20} as the model backbone. 

Different from the sparse retrieval, the dense retrieval methods concentrate on overall semantics~\citep{Zhan20,KarpukhinOMLWEC20}. For example, RepBERT~\citep{Zhan20} adopts the averaged hidden representation as query or document representation, and optimizes it with a margin loss. In contrast, DPR~\citep{KarpukhinOMLWEC20} makes the optimization easier with the \texttt{[CLS]} representation and an in-batch negative log likelihood loss. Accounting for the sub-optimality of using a single vector representation~\citep{Zhan20,KarpukhinOMLWEC20}, multi-vector methods are imposed~\citep{KhattabZ20,HumeauSLW20,LuanETC21,TangSJWZW20}. For example, ColBERT~\citep{KhattabZ20} proposes a sum-of-maxsim operation to aggregate multi-vector interactions. Similarly, PolyEncoder~\citep{HumeauSLW20} introduces complicated attention operations to achieve the same goal. On another note, as dense retrievers take the form of dual-encoders, knowledge distillation~\citep{HintonVD15} is utilized to transfer the ability of rerankers, which are usually cross-encoders, to the retrievers~\citep{TahamiGS20,RenQLZSWWW21}. Moreover, it is known that training negatives are of vital importance for dense retrievers, and various negative mining techniques have been presented~\citep{XiongXLTLBAO21,QuDLLRZDWW21,ZhanM0G0M21}. 

Despite the compelling performance of pretrained LMs, they are not tailored for dense retrieval. Therefore, quite a lot of work has been investigated to pretrain LMs for dense retrieval~\citep{LeeCT19,ChangYCYK20,LuHXKMDBLO21,GaoC21,GaoC22,Ma22,MaGZFC22}. For example, ICT~\citep{LeeCT19} constructs contrast pairs from unsupervised corpus to pretrain the retrievers. SEED~\citep{LuHXKMDBLO21} argues a weak decoder is significant for pretraining a strong encoder. In addition, Condenser~\citep{GaoC21} proposes an objective that enhances the expressiveness of the \texttt{[CLS]} representation.

While both sparse and dense methods can be compute inefficient, the efficiency bottlenecks of dense retrievers are more commonly witnessed. Thus in our work, we focus on dense retrieval and adapt \textsc{ElasticLM} to it, so as to improve not only the efficiency of dense retrieval when the number of requests is large, but also the effectiveness when the number of requests is small. 

\subsection{Reranking}

Besides the first-stage retrieval, \textsc{ElasticLM} can be flexibly extended to rerankers (e.g., PROP~\citep{MaGZFJC21b}). Usually, the reranking stage strictly follows the retrieval stage and aims to compensate the performance of the retrieval. To do so, rerankers, as mentioned above, are usually cross-encoders instead of dual-encoders as retrievers are.

Ever since the invention of LMs, they have been deployed as rerankers~\citep{Nogueira19,GaoDC21b} and instantly these LM-based rerankers are found to be not efficient enough. This invokes a hype of either decoupling query-passage interactions within rerankers~\citep{MacAvaneyN0TGF20,GaoDC20b} or distilling rerankers to smaller ones~\citep{GaoDC20,ChenHHSS21}. For exmaple, PreTTR~\citep{MacAvaneyN0TGF20} decouples the query-passage interactions at bottom layers and join the interactions at top layers, thus admitting representations of passages from bottom layers be cached. Moreover, logits from language modeling are viewed as distillation features and enable successful distillation without much performance degradation. 

Likewise, while directly applying existing LMs to reranking has already been promising, LMs tailored to reranking seem to be even more appealing~\citep{MaGZFJC21b,MaGZFLC21,MaDXZJCW21,ChenLFMFYXZM22}. For example, PROP~\citep{MaGZFJC21b} designs a representative word prediction task to enhance the modeling of inter-word correlations of LMs. HARP~\citep{MaDXZJCW21} makes hyperlinks as supervisions to augment the pretraining expressiveness. ARES~\citep{ChenLFMFYXZM22} introduces axiomatic regularization during pretraining of rerankers.

Distillation of rerankers is not new as we went through, however, pioneering studies usually focus on a static tradeoff. In our work, we attempt to extend \textsc{ElasticLM} to reranking and offer \textsc{ElasticRanker} as an alternative for expected elastic tradeoff.

\section{Compute Elasticity}

\subsection{Desiderata of Compute Elasticity}

Owing to the observation that a static tradeoff between latency and performance might not fit for all cases, we argue the necessity of bringing in compute elasticity for an elastic tradeoff. When we alter the compute of the LM elastically, the latency and performance are also changed accordingly. For example, if we decrease the compute consumed by the LM, the latency and performance would also decrease, realizing a different tradeoff. To define compute elasticity, we first list three necessary properties:

\begin{property}[online]
Compute is online if it could vary on-the-fly. Compute can be either online or offline, i.e., it is binary.
\end{property}


\begin{property}[scalable] 
Compute is scalable if it could be scaled up and down. 
\end{property}


\begin{property}[controllable]
Compute is controllable if it could be scaled to a defined value contingent on a specific context.
\end{property}

In summary, we define a compute elasticity to have the above three properties:
\begin{definition}[compute elasticity]
    Compute elasticity denotes online,  scalable, and controllable compute.
\end{definition}

The context of previous state of the art is mainly the input, e.g., EarlyExitBERT~\citep{LiuSHWWZJCHQ22}. Our work concentrates on a distinguished context, i.e., the number of requests, since the tradeoff is in fact expected to vary along varying request stream. A glimpse of arts are recapped in Table~\ref{tab:1}. They can partly achieve our expectations but not completely, hinting the elastic tradeoff is nontrivial.


\begin{table}[t]
    \centering
    \caption{The glimpse of arts.}
    \begin{adjustbox}{width=0.65\textwidth,center}
    \begin{tabular}{lccc}
    \toprule
        \textbf{Method} & \textbf{Online} & \textbf{Scalable} & \makecell[c]{\textbf{Controllable}\\\textbf{w.r.t. Request Stream}} \\
    \midrule
        TinyBERT~\citep{JiaoYSJCL0L20} & \xmrk & \xmrk & \xmrk \\
        DynaBERT~\citep{HouHSJCL20} & \xmrk  & \cmrk & \xmrk \\
        EarlyExitBERT~\citep{LiuSHWWZJCHQ22} & \cmrk & \cmrk & \xmrk \\
    \midrule
        Compute Elasticity & \cmrk & \cmrk & \cmrk \\
    \bottomrule
    \end{tabular}
    \end{adjustbox}
    \label{tab:1}
\end{table}

\subsection{Formulation of Compute Elasticity}

We hereby formulate our core problem. Traditionally, given a teacher LM with a latency-performance tradeoff $\bigl\{(l^{\sf T},p^{\sf T})\bigl\}$ associated with a structure $a^{\sf T}$, knowledge distillation gives a student LM with a latency-performance tradeoff $\bigl\{(l^{\sf S},p^{\sf S})\bigl\}$ associated with a structure $a^{\sf S}$. Here, $l$ and $p$ respectively indicate latency and performance, and  $l^{\sf S}\leq l^{\sf T}$ and $p^{\sf S}\leq p^{\sf T}$ strictly hold. With slight abuse of notation, we annotate $\vert a \vert$ as the scale of a structure; therefore $\vert a^{\sf S} \vert \leq \vert a^{\sf T} \vert$. Unlike the traditional case, \textsc{ElasticLM} should be distilled with a series of latency-performance tradeoffs $\bigl\{(l^{\sf E}_{k},p^{\sf E}_{k})\bigl\}_{k}$ for compute elasticity; and corresponding structures are denoted as $\bigl\{a^{\sf E}_{k}\bigl\}_{k}$. Herein, we treat BERT~\citep{DevlinCLT19} as the teacher LM. 

\section{Elastic Language Model}

\begin{figure}[t]
    \centering
    \includegraphics[width=0.97\textwidth]{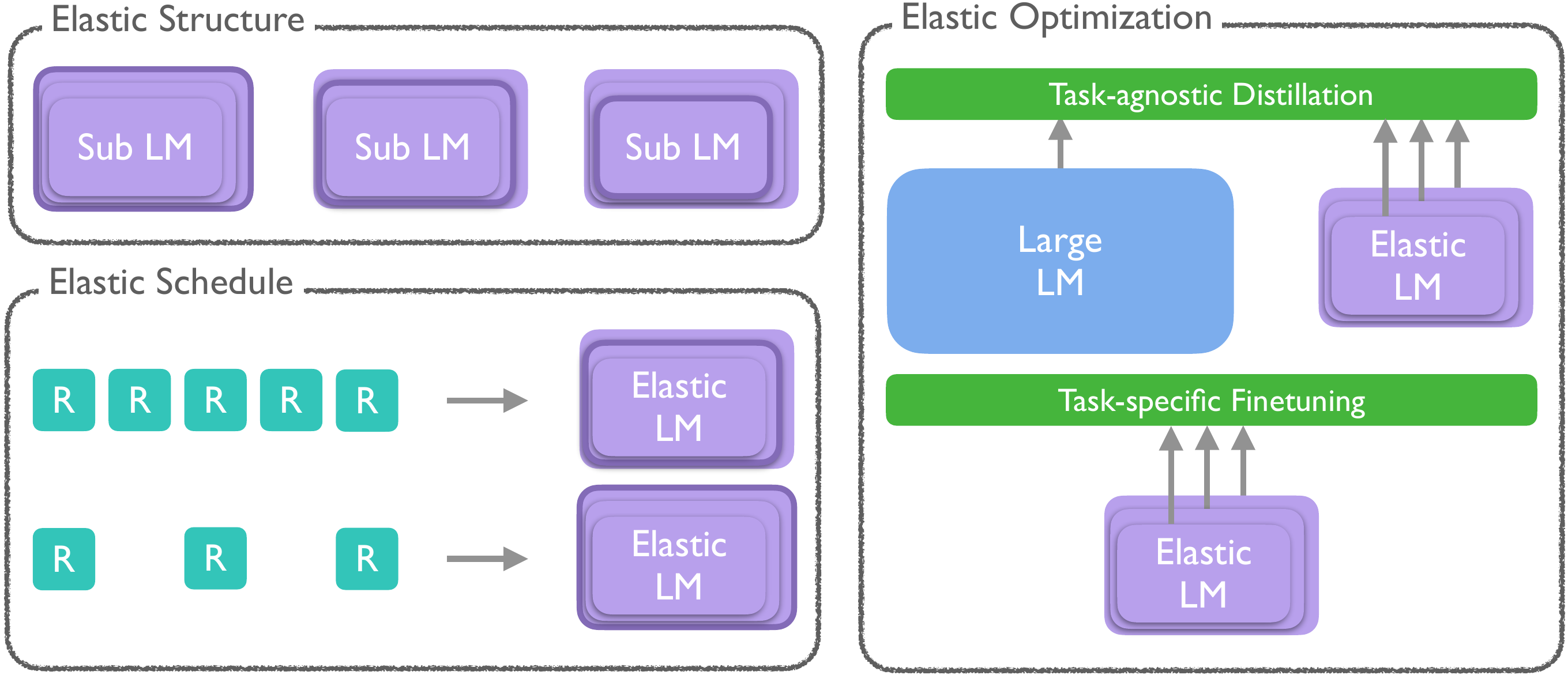}
    \caption{The overview of \textsc{ElasticLM}. Elastic structure enables \textsc{ElasticLM} with compute elasticity; elastic optimization learns \textsc{ElasticLM} under compute elasticity; and elastic schedule drives \textsc{ElasticLM} for compute elasticity.}
    \label{fig:2}
\end{figure}

\textsc{ElasticLM} involves three major designs: an elastic structure, an elastic optimization, and an elastic schedule. An overview of \textsc{ElasticLM} is given in Figure~\ref{fig:2}.

\subsection{Elastic Structure}

\paragraph{Atomicity in Transformer}

An encoder-only LM (e.g., BERT) consists of a stack of transformer layers~\citep{VaswaniSPUJGKP17}, each of which further includes a multihead self-attention block (MHA) and a feedforward network block (FFN). Concretely, given an $n$-length sequence of $d$-dimensional input vectors $\mathbf{X}\in\mathbb{R}^{n\times d}$, the output of the MHA block with $A$ independent heads can be represented as:
\begin{equation}
    \textrm{MHA}(\mathbf{X})=\sum_{j=1}^{A}\textrm{Attn}(\mathbf{X};\mathbf{W}^{\sf Q}_{j},\mathbf{W}^{\sf K}_{j})\mathbf{X}\mathbf{W}^{\sf V}_{j}\mathbf{W}^{\sf O}_{j},
\end{equation}
where the $j$-th head is parameterized by $\mathbf{W}^{\sf Q}_{j}$, $\mathbf{W}^{\sf K}_{j}$, $\mathbf{W}^{\sf V}_{j}\in\mathbb{R}^{d\times d^{\sf A}}$, and $\mathbf{W}^{\sf O}_{j}\in\mathbb{R}^{d^{\sf A}\times d}$. On the other hand, the output of the FFN block is shown as: 
\begin{equation}
    \textrm{FFN}(\mathbf{X})=\textrm{GELU}(\mathbf{X}\mathbf{W}^{\sf I})\mathbf{W}^{\sf O}=\sum_{j=1}^{I}  \textrm{GELU}(\mathbf{X} \mathbf{W}^{\sf I}_{\cdot,j})\mathbf{W}^{\sf O}_{j,\cdot}
\end{equation}
where two fully-connected layers are parameterized by $\mathbf{W}^{\sf I}\in\mathbb{R}^{d\times d^{\sf I}}$ and $\mathbf{W}^{\sf O}\in\mathbb{R}^{d^{\sf I}\times d}$ respectively. Details such as biases and normalizations of a transformer layer, are omitted for brevity.

In certain sense, attention heads in an MHA block and intermediate neurons in an FFN block are atomic modules and could compose to compound modules, as defined below:

\begin{definition}[atomic and compound modules]
A compute module $g_j$ is an atomic module if it is the minimum independent component.
Let $H$ be the number of all atomic modules. 
Accordingly, a compute module $f^\prime$ is a compound module if it is a sum of available atomic modules: $f^{\prime}(\mathbf{X})=\sum_{j=1}^{H^{\prime}}g_j(\mathbf{X})$. Here, $H^{\prime}\leq H$ is the number of available atomic modules. Particularly, $f$ is also a special compound module that $f(\mathbf{X})=\sum_{j=1}^{H}g_j(\mathbf{X})$.
\end{definition}

\paragraph{Atomic and Compound Structure}

Thanks to that LMs are essentially stacked transformers, we can view an attention head or an intermediate neuron as an atomic structure, while any compound structure could be an aggregation of these atomic structures.

A naive way to enable \textsc{ElasticLM} with compute elasticity is to maintain a range of different structures, with distinct latency-performance tradeoffs, i.e., $\bigcap_{k} a^{\sf E}_{k}=\varnothing$. However, doing so would lead to unexpectedly huge memory footprint.
Instead, based on the atomic formation, a \textsc{SubLM} can exactly correspond to a compound structure and an elastic structure thus arranges an ensemble of compound structures in a way that $a^{\sf E}_{k}\subseteq a^{\sf E}_{h}$ and $\vert a^{\sf E}_{k} \vert\leq \vert a^{\sf E}_{h}\vert$ exist. Here, an equivalent interpretation is that: $a^{\sf E}_{k}$ is already a compound structure, and $a^{\sf E}_{h}$ is obtained by adding a few atomic structures.

Built upon the elastic structure, {\rm \textsc{ElasticLM}} in fact can only occupy memory that is equal to that of {\rm \textsc{SubLM}} owning $\max_{k} a^{\sf E}_{k}$ but at the same time is an ensemble of {\rm \textsc{SubLM}}s. Each {\rm \textsc{SubLM}} has a structure that is a subset of that of its larger counterpart. 

\paragraph{Pruned Compound Structure}

\begin{figure}[t]
    \centering
    \includegraphics[width=0.65\textwidth]{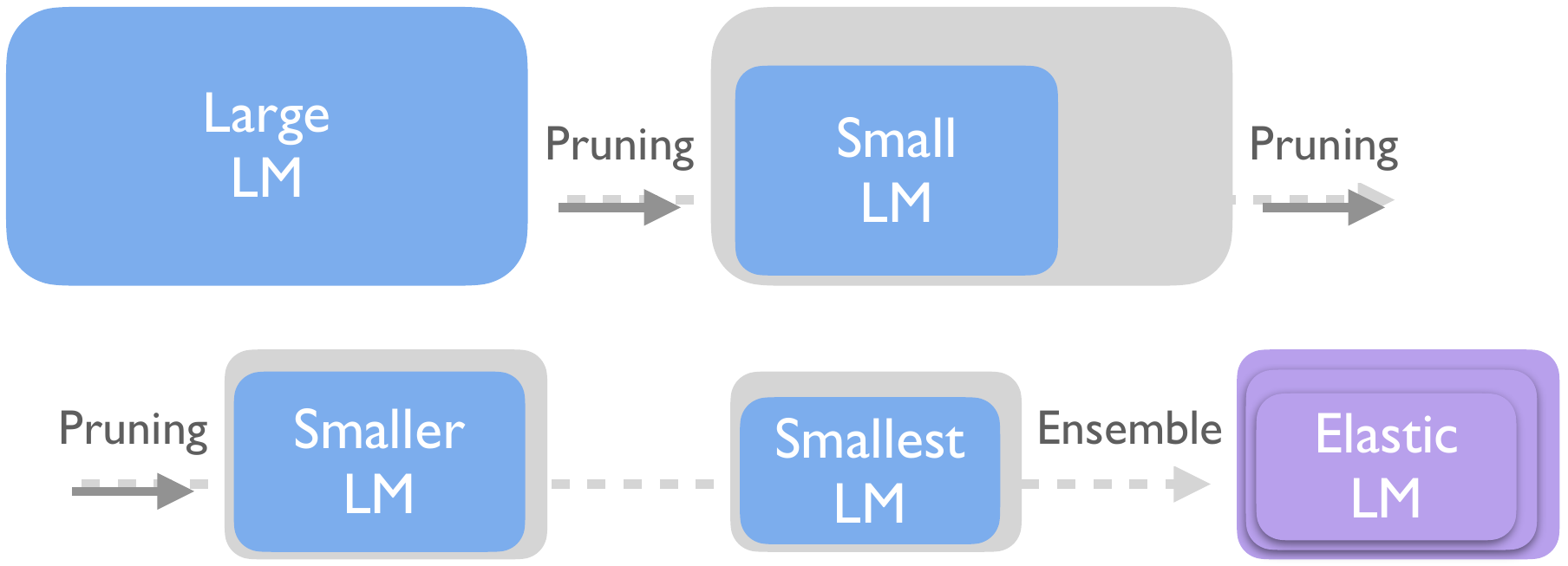}
    \caption{The illustration of elastic structure. Incrementally pruning a large teacher LM generates a set of smaller structures for SubLMs, and assembling these smaller structures in a parameter-sharing fashion results in the elastic structure for \textsc{ElasticLM}.}
    \label{fig:3}
\end{figure}

For \textsc{ElasticLM}, we can assign it a structure initiation (i.e., $\max_{k} a^{\sf E}_{k}$) by pruning the teacher to a corresponding level of preserved amount of parameters. Subsequently, we can get smaller structures (i.e., $a^{\sf E}_{k} < \max_{h} a^{\sf E}_{h}$) of \textsc{SubLM}s by incrementally pruning the teacher. Due to the atomicity in these structures, these \textsc{SubLM}s are assembled into \textsc{ElasticLM} via parameter sharing. An illustration of elastic structure is given in Figure~\ref{fig:3}.

To uncover these \textsc{SubLM}s, we mainly prune the attention heads of MHA blocks and intermediate neurons of FFN blocks from the teacher following parameter expressive score~\citep{MolchanovTKAK17}. Following the literature on structured pruning in a post-training style~\citep{MichelLN19,HouHSJCL20}, we attach a set of variables $\xi_{j}$ and $\nu$ to the attention heads and the intermediate neurons, to record the parameter sensitivities through accumulated absolute gradients, as shown below:
\begin{equation}
    {\rm MHA}^{\circ}(\mathbf{X})
    =\sum_{j=1}^{A}\xi_{j}{\rm Attn}(\mathbf{X},\mathbf{W}^{\sf Q}_{j},\mathbf{W}^{\sf K}_{j},\mathbf{W}^{\sf V}_{j})\mathbf{W}^{\sf O}_{j},
\end{equation}
\begin{equation}
    {\rm FFN}^{\circ}(\mathbf{X})={\rm GELU}(\mathbf{X}\mathbf{W}^{\sf I}){\rm diag}(\nu)\mathbf{W}^{\sf O},
\end{equation}
where $\xi_{j}\equiv 1$ and $\nu\equiv\mathbf{1}^{d^{\sf I}}$. We set the values of the $\xi_{j}$ and $\nu$ to ones to ensure the functionalities of corresponding heads and neurons are retained. 

\paragraph{Expressive Score}

A higher expressive score indicates that the corresponding parameter has bigger contribution towards the loss. Specifically, the expressive scores of the attention heads in MHA and the intermediate neurons in FFN are depicted as:
\begin{equation}
\begin{aligned}
    &\mathbb{I}^{\sf head}_{j}=\mathbb{E}_{\mathcal{D}^{\sf MLM}}\left|\frac{\partial\mathcal{L}^{\sf MLM}}{\partial\xi_{j}}\right|, \\
    &\mathbb{I}^{\sf neuron}=\mathbb{E}_{\mathcal{D}^{\sf MLM}}\left|\frac{\partial\mathcal{L}^{\sf MLM}}{\partial {\rm diag}(\nu)}\right|,
\end{aligned}
\end{equation}
where $\mathcal{D}^{\sf MLM}$ is the data distribution for masked language modeling (e.g., Wikipedia), and $\mathbb{E}$ represents expectation.

\begin{figure}[t]
    \centering
    \includegraphics[width=0.47\textwidth]{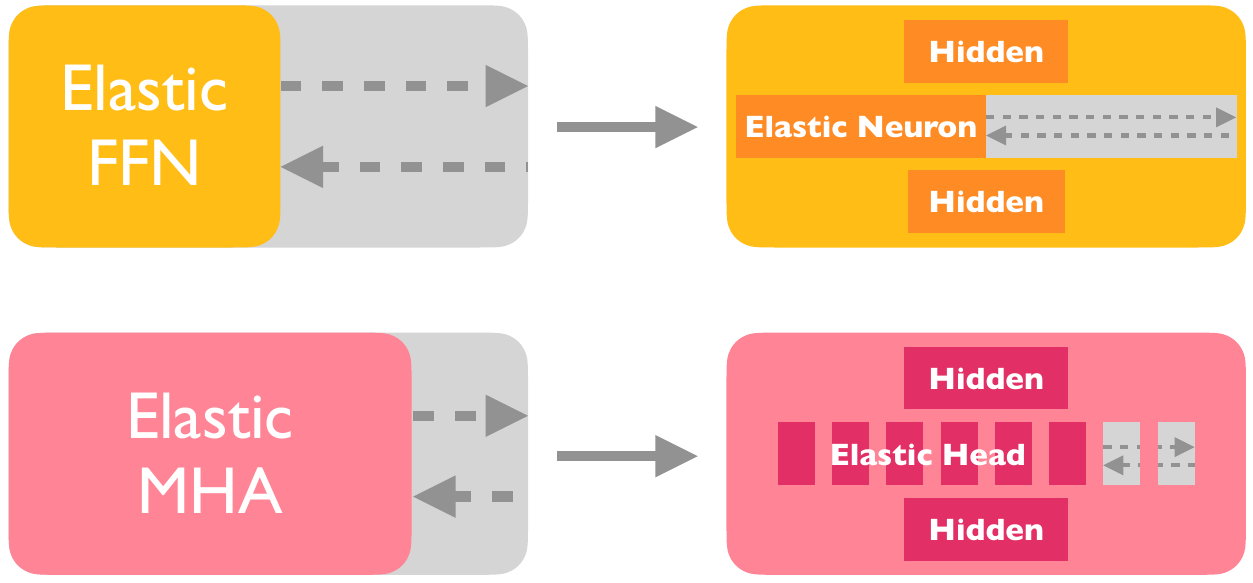}
    \caption{The depiction of how MHA and FFN blocks behave within \textsc{ElasticLM}. The attention heads in each MHA block and the intermediate neurons in each FFN block shall be adjusted elastically.}
    \label{fig:4}
\end{figure}

Referring to~\citet{MolchanovTKAK17}, we normalize the expressive scores with $\ell_{2}$ norm. The parameters in the teacher with adequately low scores can be pruned such that MHA and FFN blocks in \textsc{ElasticLM} behave akin to what is depicted in Figure~\ref{fig:4}.

\paragraph{Intuition behind Expressive Score}

This implementation is mathematically equivalent to the prevalent first-order Taylor expansion of the absolute loss variation between before and after removing a module (i.e., a head or a neuron). Taking the $j$-th attention head as an example, its parameter sensitivity can be written as:
\begin{equation}
\begin{aligned}
    &\left|\frac{\partial\mathcal{L}^{\sf MLM}}{\partial\xi_{j}}\right|=\left|\frac{\partial\mathcal{L}^{\sf MLM}}{\partial\xi_{j}\mathbf{O}_{j}}\frac{\partial\xi_{j}\mathbf{O}_{j}}{\partial\xi_{j}}\right|=\left|\frac{\partial\mathcal{L}^{\sf MLM}}{\partial\mathbf{O}_{j}}\mathbf{O}_{j}\right|\\
    &\approx\left|(\mathcal{L}^{\sf MLM}_{\mathbf{0}}+\frac{\partial\mathcal{L}^{\sf MLM}}{\partial\mathbf{O}_{j}}(\mathbf{O}_{j}-\mathbf{0})+\mathbf{r})-\mathcal{L}^{\sf MLM}_{\mathbf{0}}\right|\\
    &=\left|\mathcal{L}^{\sf MLM}-\mathcal{L}^{\sf MLM}_{\mathbf{0}}\right|,
\end{aligned}
\end{equation}
where $\mathcal{L}^{\sf MLM}$ stands for the pretraining objective of the teacher (i.e., masked language modeling in our case), and $\mathbf{O}_{j}$ is utilized for $j$-th attention head output. $\mathcal{L}^{\sf MLM}_{\mathbf{0}}$ actually means $\mathcal{L}^{\sf MLM}|_{\mathbf{O}_{j}=\mathbf{0}}$, and $\mathbf{r}$ represents residuals in Taylor expansion.

\subsection{Elastic Optimization}

A natural obligation after the elastic structure is to optimize the \textsc{ElasticLM}, for which we design an elastic optimization.

The elastic optimization makes an elastic structure converge, so that $p^{\sf E}_{k}$ associated with each structure $a^{\sf E}_{k}$ is guaranteed, by decomposing each optimization step as a traverse of all structures in the ensemble. In the elastic optimization, {\rm \textsc{ElasticLM}} is optimized  essentially through enumerating and optimizing each {\rm \textsc{SubLM}} at every optimization step.

We distil \textsc{ElasticLM} from the teacher in a task-agnostic manner due to its performance superiority~\citep{Turc19,WangW0B0020} for small LMs. Simply put, \textsc{ElasticLM} is pretrained via distillation from the teacher. We adopt a state-of-the-art relation alignment distillation objective~\citep{WangBHDW21} $\mathcal{L}^{\sf ALN}$ as below:
\begin{equation}
\begin{aligned}
    \mathcal{L}^{\sf ALN}=\mathbb{E}_{\mathbf{X}\sim\mathcal{D}^{\sf ALN}}\sum_{j=1}^{R} 
    &\textrm{KL}(\textrm{Reln}(\mathbf{X};{}^{\mathcal{T}}\mathbf{W}^{\sf Q}_{j}),\textrm{Reln}(\mathbf{X};{}^{\mathcal{S}}\mathbf{W}^{\sf Q}_{j})) \\
    &  + \textrm{KL}(\textrm{Reln}(\mathbf{X};{}^{\mathcal{T}}\mathbf{W}^{\sf K}_{j}),\textrm{Reln}(\mathbf{X};{}^{\mathcal{S}}\mathbf{W}^{\sf K}_{j})) \\
    &+\textrm{KL}(\textrm{Reln}(\mathbf{X};{}^{\mathcal{T}}\mathbf{W}^{\sf V}_{j}),\textrm{Reln}(\mathbf{X};{}^{\mathcal{S}}\mathbf{W}^{\sf V}_{j})),
\end{aligned}
\end{equation}
\begin{equation}
    \textrm{Reln}(\mathbf{X};{}^{\mathcal{T}}\mathbf{W}^{\sf Q}_{j})=\textrm{softmax}(\mathbf{X}{}^\mathcal{T}\mathbf{W}^{\sf Q}_{j}{}^\mathcal{T}\mathbf{W}^{\sf Q\top}_{j}\mathbf{X}^{\top}/d^{\sf R}),
\end{equation}
where $\mathcal{D}^{\sf ALN}$ is the data distribution for task-agnostic distillation (usually the same as that for pretraining the teacher), and KL stands for Kullback-Leibler divergence. Necessarily, relation heads are derived by merging the original $A$ attention heads and then splitting them to $R$ heads. ${}^{\mathcal{T}}\mathbf{W}^{\sf Q}_{j}$ is the redistributed query parameter of the $j$-th relation head within totally $R$ heads from the last MHA block of the LM, likewise ${}^{\mathcal{T}}\mathbf{W}^{\sf K}_{j}$ and ${}^{\mathcal{T}}\mathbf{W}^{\sf V}_{j}$ are the key and value parameters. Since some heads are pruned for a few \textsc{SubLM}s, an additional MHA block is built upon \textsc{ElasticLM} to serve as the last MHA block. \textsc{ElasticLM} can be then finetuned on any task-specific data distribution  $\mathcal{D}^{\sf TSK}$ and learning objective $\mathcal{L}^{\sf TSK}$ in a way similar to how it is elastically pretrained as in Algorithm~\ref{alg:1}.

\begin{figure}[t]
\begin{algorithm}[H]
    \definecolor{codeblue}{rgb}{0.25,0.5,0.5}
    \lstset{
      backgroundcolor=\color{white},
      basicstyle=\fontsize{10.2pt}{10.2pt}\ttfamily\selectfont,
      columns=fullflexible,
      breaklines=true,
      captionpos=b,
      commentstyle=\fontsize{10.2pt}{10.2pt}\color{codeblue},
      keywordstyle=\fontsize{10.2pt}{10.2pt},
    }
    \begin{lstlisting}[language=python]
    # lm_t, lm_s: teacher LM and student LM (ElasticLM).
    # attn: additional MHA layer for student LM.
    # submap: key-value map to notify which params to drop for sub student LM.
    
    for x in dataloader:  # load a minibatch x.
        r_t = lm_t(x).relations  # relations of teacher LM.
        r_t = r_t.detach() # no gradients to teacher LM.        
        
        for k in submap:
            # index params to realize sub student LM,
            # i.e., mark the params belonging to the sub LM so that they would be used during forward computation.
            index(lm_s, submap[k])
            
            h_s = lm_s(x).hiddens # hiddens of sub student LM.
            r_s = attn(h_s).relations # relations of sub student LM.
            
            # loss is computed and gradients are accumulated.
            loss = KLDivLoss(r_t, r_s) / len(submap)
            loss.backward()
            
        # update student LM with accumulated gradients.
        update(lm_s.params)
    \end{lstlisting}
    \caption{PyTorch-like pseudocode of elastic optimization.}
    \label{alg:1}
\end{algorithm}
\end{figure}

\subsection{Elastic Schedule}

\begin{figure}[b]
    \centering
    \includegraphics[width=0.52\textwidth]{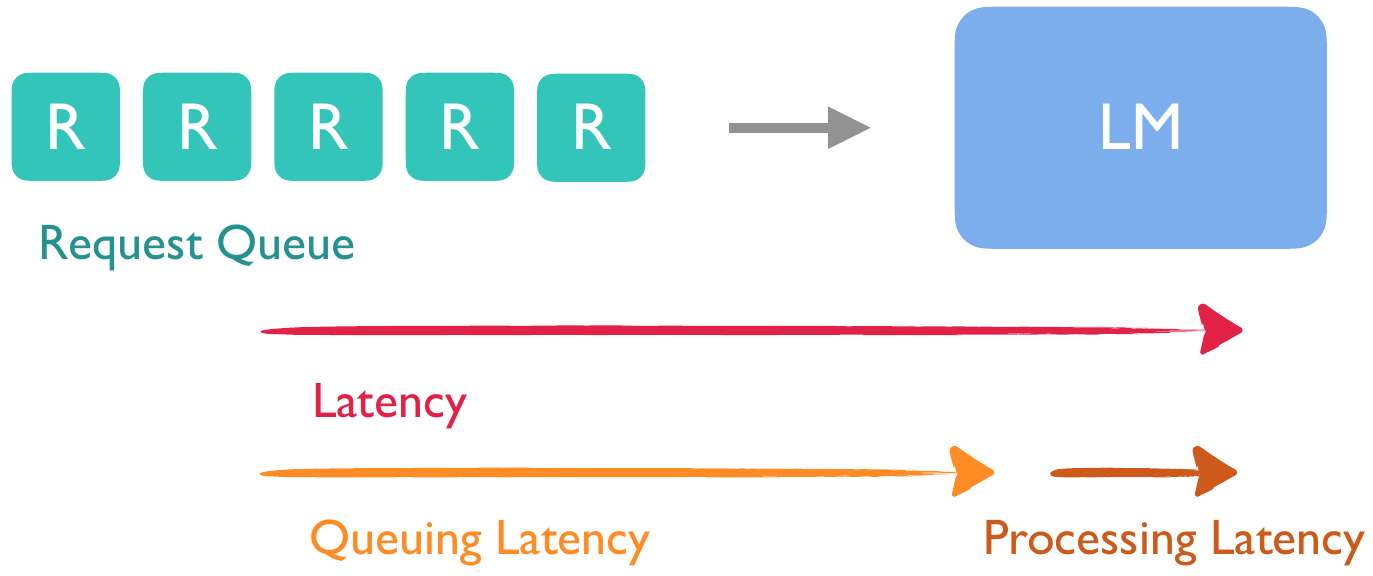}
    \caption{The sketch of latency decomposition.}
    \label{fig:5}
\end{figure}

The latency can be decomposed into the \textbf{p}rocessing latency $t^{\sf p}$ and \textbf{q}ueuing latency $t^{\sf q}$, and the queuing time is roughly a multiple of the processing time, where the multiplier is the queue size $g$. Accordingly, the latency of any \textsc{SubLM} with structure $a^{\sf E}_{k}$ can be estimated as $l^{\sf E}_{k}=t^{\sf p}_{k}+t^{\sf q}_{k}\approx (g+1)\cdot t^{\sf p}_{k}$. A sketch of the latency decomposition is shown in Figure~\ref{fig:5}.

Therefore, given a latency constraint $T$, an elastic schedule should opt to \textsc{SubLM}s satisfying the constraint by reinforcing the inequality $(g+1)\cdot t^{\sf p}_{k} \leq T$, i.e., $g\leq T/t^{\sf p}_{k}-1$. 

The elastic schedule manages an elastic structure by referring to the instant queue size $g$ and the constant latency constraint $T$, while maximally retaining the performance. The elastic schedule determines that the {\rm \textsc{SubLM}} with structure $a^{\sf E}_{k}$ should not be used when the instant queue size $g$ is larger than $T/t^{\sf p}_{k}-1$ and smaller {\rm \textsc{SubLM}}s should be used instead. Further, to maximally retain the performance, the {\rm \textsc{SubLM}} with the highest performance should be selected among these smaller {\rm \textsc{SubLM}}s. Reversely, when the instant queue size $g$ otherwise is smaller than $T/t^{\sf p}_{k}-1$, larger {\rm \textsc{SubLM}}s should be shifted to.

For distinguished latency constraints, different elastic schedules are automatically calibrated since no other information is needed.

\subsection{Adaptation to Information Retrieval}

\begin{figure}[b]
    \centering
    \includegraphics[width=0.45\textwidth]{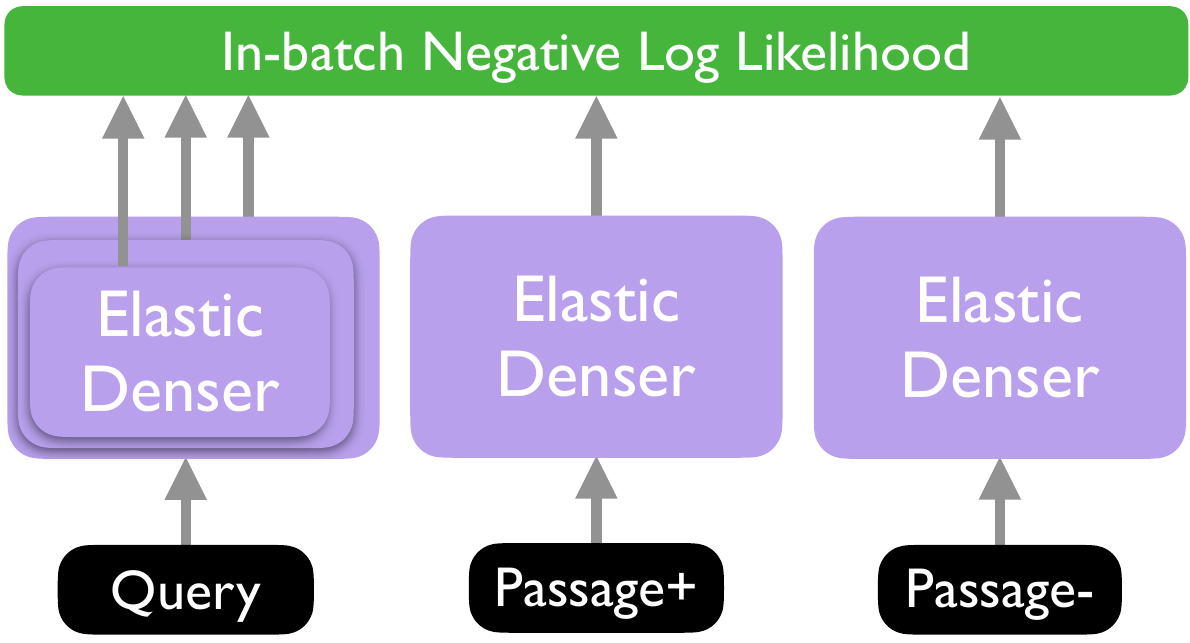}
    \caption{The description of \textsc{ElasticDenser}. During task-specific finetuning, the passage encoder consumes only the largest structure, while the query encoder traverses all structures.}
    \label{fig:6}
\end{figure}

Since IR scenarios are sensitive to latency variation, we adapt \textsc{ElasticLM} to dense retrieval.

Provided that \textsc{ElasticLM} is task-agnostic and Condenser~\citep{GaoC21} and PROP~\citep{MaGZFJC21b} are both continuously pretrained BERTs but tailored separately for dense retrieval and reranking, \textsc{ElasticLM} can be flexibly adapted to dense retrieval and reranking by treating Condenser or PROP instead of BERT as the teacher LM. Accordingly, we term \textsc{ElasticLM} with Condenser being the teacher as \textsc{ElasticDenser} and that with PROP being the teacher as \textsc{ElasticRanker}.

We preserve the elastic structure attained from BERT for simplicity, but replace BERT with Condenser or PROP during elastic optimization for performance boost. Apparently, the task-specific objective should take a form that is suitable for dense retrieval or reranking. 

Inspired by DPR~\citep{KarpukhinOMLWEC20}, we instantiate $\mathcal{L}^{\sf TSK}$ for \textsc{ElasticDenser} to in-batch negative log likelihood as below:
\begin{equation}
    \mathcal{L}^{\sf TSK}=\mathbb{E}_{(q^{}_{i},p^{+}_{i},p^{-}_{i,1},\cdots,p^{-}_{i,m-1})\sim\mathcal{D}^{\sf TSK}} -\mathrm{log}\frac{e^{\mathrm{sim}(q^{}_{i},p^{+}_{i})}}{e^{\mathrm{sim}(q^{}_{i},p^{+}_{i})}+\sum_{l=1}^{m-1} e^{\mathrm{sim}(q^{}_{i},p^{-}_{i,l})}}
    \label{eqn:9}
\end{equation}
\begin{equation}
    \mathrm{sim}(q,p)=a^{\sf E}_{k}(q)_{\tt [CLS]}^{\top}\{\max_{k} a^{\sf E}_{k}\}(p)_{\tt [CLS]}^{}
\end{equation}
where $(q^{}_{i},p^{+}_{i},p^{-}_{i,1},\cdots,p^{-}_{i,m-1})$ denotes that one question/query has one relevant positive passages and $m-1$ irrelevant negative passages. The negatives are comprised of both in-batch and hard negatives retrieved by BM25, and the similarity between the question/query and the passage is computed with a simple dot product over their \texttt{[CLS]} representations in an shared encoder (i.e., \textsc{ElasticDenser}) like what has been done in Condenser rather than two separate encoders in DPR. 

Another consideration is that the passage encoder would be always kept as the largest structure. On the one hand, since passages are usually encoded offline, this behaviour can hold the performance as much as possible without affecting the latency. On the other hand, more importantly, there will otherwise be multiple cached embeddings for one passage due to the elastic structure, giving rise to unreasonable memory consumption. Figure~\ref{fig:6} offers a description of \textsc{ElasticDenser} during task-specific finetuning.

Following localized contrastive estimation (LCE,~\citep{GaoDC21b}), we instantiate $\mathcal{L}^{\sf TSK}$ for \textsc{ElasticRanker} to localized negative loglikelihood, whose basic form strictly follows Equation~\ref{eqn:9} and only differs in:
\begin{equation}
    \mathrm{sim}(q,p)=\mathbf{w}^{\top}a^{\sf E}_{k}(\mathrm{concat}(q,p))_{\tt [CLS]}
\end{equation}
Here, the similarity between the question/query and the passage is computed with a linear mapping $\mathbf{w}$ over the \texttt{[CLS]} representations of concatenated query-passage inputs. Since the reranker is usually plugged to the retriever as the second stage, the negatives $(p^{-}_{i,1},\cdots,p^{-}_{i,m-1})$ are constructed by sampling from false positives lying in the top ranked passages of a retriever so that the reranker is optimized towards the retriever. And this is also the reason why such contrastive estimation is called localized. 

\section{Experiments}

\subsection{Task-agnostic Distillation}

\begin{table}[b]
    \centering
    \caption{The data statistics and corresponding metrics.}
    \begin{adjustbox}{width=0.85\textwidth,center}
    \begin{tabular}{lrrrc}
    \toprule
        \textbf{Dataset} & \textbf{\#Train exam.} & \textbf{\#Dev exam.} & \textbf{\#Test exam.} & \textbf{Metric} \\
    \midrule
        \multicolumn{5}{c}{\textit{Task-agnostic Distillation}} \\
    \midrule
        Wikipedia & 35M & - & - & - \\
    \midrule
        \multicolumn{5}{c}{\textit{Language Understanding}} \\
    \midrule
        SST-2 & 67K & 0.9K & - & Accuracy \\
        MRPC & 3.7K & 0.4K & - & F1 \\
        STS-B & 7K & 1.5K & - & Spearman Correlation \\
        QQP & 364K & 40K & - & F1 \\
        MNLI-m/mm & 393K & 20K & - & Accuracy \\
        QNLI & 105K & 5.5K & - & Accuracy \\
        RTE & 2.5K & 0.3K & - & Accuracy \\
    \midrule
        \multicolumn{5}{c}{\textit{Dense Retrieval}} \\
    \midrule
        NQ & 58.9K & 8.8K & 3.6K & Recall@\{5,20,100\} \\
        TriviaQA & 60.4K & 8.8K & 11.3K & Recall@\{5,20,100\} \\
        MC Psg & 532.8K & 7K & - & MRR@\{10,100\} \\
    \midrule
        \multicolumn{5}{c}{\textit{Reranking}} \\
    \midrule
        MC Doc & 367K & 5.2K & - & MRR@\{10,100\} \\
    \bottomrule
    \end{tabular}
    \end{adjustbox}
    \label{tab:101}
\end{table}

\paragraph{Data}

Following previous studies in task-agnostic distillation~\cite{JiaoYSJCL0L20,WangW0B0020}, we use pretraining data to distil \textsc{ElasticLM}, \textsc{ElasticDenser}, and \textsc{ElasticRanker}. While BERT is pretrained on Wikipedia and BookCorpus~\citep{ZhuKZSUTF15}, we find that using only Wikipedia for task-agnostic distillation is fairly enough. After preprocessing, the data contains around 35M examples.

\paragraph{Implementation}

The pruning procedure (i.e., obtaining elastic structure) is conducted on a random subset of Wikipedia. As any model structure $a^{\sf E}_{k}$ exactly corresponds to a preserving level during pruning, we assign seven preserving levels \{50,40,30,20,15,10,5\}\% and  get an ensemble of seven structures, where the highest preserving level corresponds to $\max_{k} a^{\sf E}_{k}$.  The teachers are BERT\textsubscript{\sf base}, Condenser\textsubscript{\sf base}, PROP\textsubscript{\sf base} for \textsc{ElasticLM}, \textsc{ElasticDenser}, \textsc{ElasticRanker}, respectively.

The pretraining (i.e., the first step of achieving elastic optimization) of \textsc{ElasticLM}, \textsc{ElasticDenser}, and \textsc{ElasticRanker} is not much different except the opted teacher. The pretraining procedure is conducted on eight Nvidia A100 GPUs. The maximum sequence length is limited to 128 for both \textsc{ElasticLM} and \textsc{ElasticDenser} while 512 for \textsubscript{ElasticRanker} according to different concerned tasks. The optimizer in use is AdamW. The batch size is set to 128 per device, which actually results in 128*8=1024 in our case. The learning rate is set to 3e-4 and the weight decay is set to 1e-2. The number of training epochs is set to 5 with the proportion of warmup epochs being 1e-2. The maximum gradient norm is limited to 5. The number of relation heads (i.e., $R$) is set to 32. Our codebase is built upon PyTorch,\footnote{\url{https://github.com/pytorch/pytorch}} and Transformers.\footnote{\url{https://github.com/huggingface/transformers}}

The data and implementation can be separately summarized in Table~\ref{tab:101} and Table~\ref{tab:102}.

\begin{table}[t]
    \centering
    \caption{The hyperparameters for both task-agnostic distillation and task-specific finetuning. The search grids are indicated with \{ \} and the potential variables are indicated with /.}
    \begin{adjustbox}{width=\textwidth,center}
    \begin{tabular}{lcccccc}
      \toprule
        \textbf{Hyperparameter} & \makecell[c]{\textbf{Task-agnostic}\\\textbf{Distillation}} & \makecell[c]{\textbf{\textsc{ElasticLM}}\\\textbf{GLUE}} & \makecell[c]{\\\textbf{NQ}} & \makecell[c]{\textbf{\textsc{ElasticDenser}}\\\textbf{Trivia}} & \makecell[c]{\\\textbf{MC Psg}} & \makecell[c]{\textbf{\textsc{ElasticRanker}}\\\textbf{MC Doc}} \\
      \midrule
        Batch size & 1024 & \{16,32\} & \multicolumn{2}{c}{128} & 8 & 4 \\
        Optimizer & AdamW & AdamW & \multicolumn{2}{c}{AdamW} & AdamW & AdamW \\
        Learning rate & 3e-4 & \{1e-5,2e-5,3e-5\} & \multicolumn{2}{c}{1e-5} & 5e-6 & 1e-5 \\
        Training epochs & 5 & 10 & \multicolumn{2}{c}{40} & 3 & 2 \\
        Warmup proportion & 0.01 & 0.1 & \multicolumn{2}{c}{0.1} & 0.1 & 0.1 \\
        Weight decay & 0.01 & 0.01 & \multicolumn{2}{c}{0.0} & 0.0 & 0.01 \\
        Gradient norm & 5.0 & 5.0 & \multicolumn{2}{c}{-} & - & - \\
        Maximum length & 128/512 & 64/128 & \multicolumn{2}{c}{32/156} & 16/128 & 512/256 \\
        Relation heads & 32 & - & \multicolumn{2}{c}{-} & - & - \\
        Negative samples & - & - & \multicolumn{2}{c}{1} & 7 & 7 \\
      \bottomrule
    \end{tabular}
    \end{adjustbox}
    \label{tab:102}
\end{table}



\subsection{Offline Evaluation}

\paragraph{Data \& Metrics}

Following the common practice in both pretraining and task-agnostic distillation~\citep{DevlinCLT19,JiaoYSJCL0L20}, we use a language understanding benchmark GLUE~\citep{WangSMHLB19} for offline evaluation of \textsc{ElasticLM}. The GLUE benchmark originally consists of two sequence classification tasks, i.e., SST-2~\citep{SocherPWCMNP13}, CoLA~\citep{WarstadtSB19}, and seven sequence-pair classification tasks, i.e., MRPC~\citep{DolanB05}, STS-B~\citep{CerDALS17}, QQP, MNLI~\citep{WilliamsNB18}, QNLI~\citep{RajpurkarZLL16}, RTE~\citep{BentivogliCDG11}, WNLI~\citep{LevesqueDM12}. We exclude WNLI and CoLA due to the evaluation inconsistency (in other words, small LMs get dramatically worse results while large LMs get much better ones as found out in~\citep{XiaZC22}) and use the remaining tasks. Following BERT~\citep{DevlinCLT19}, we report Accuracy (Acc) on SST-2, MNLI, QNLI and RTE; Spearman Correlation scores (SpCorr) on STS-B; and F1 on MRPC, QQP and CoNLL. Average score over tasks from GLUE (GLUE Score) is additionally computed. Results on development sets are reported.

Furthermore, following the pioneering work in dense retrieval~\citep{KarpukhinOMLWEC20,GaoC21}, we use two open-domain question answering datasets: Natural Question (NQ)~\citep{KwiatkowskiPRCP19} and Trivia QA (Trivia)~\citep{JoshiCWZ17}, and a passage ranking dataset MS MARCO Passage Ranking (MC Psg)~\citep{NguyenRSGTMD16} for offline evaluation of \textsc{ElasticDenser}. Both NQ and Trivia are preprocessed according to DPR~\citep{KarpukhinOMLWEC20}. We report Recall scores of top 5, 20, and 100 retrieved passages (R@\{5,20,100\}) on NQ and Trivia; and Mean Reciprocal Rank scores at top 10 retrieved passages (MRR@10) on MC Psg. Results on test sets are reported on NQ and Trivia, and results on development sets are reported on MC Psg.

Following the literature in reranking~\citep{GaoDC21b}, we mainly use a document ranking dataset MS MARCO Document Ranking (MC Doc) for offline evalution of \textsc{ElasticRanker}. And we additionally plug \textsc{ElasticRanker} to \textsc{ElasticDenser} for an evaluation of \textsc{ElasticRanker} on MC Psg. We report Mean Reciprocal Rank scores at top 10 and 100 retrieved documents (MRR@\{10,100\}) on MC Doc and similarly on MC Psg. Results on development sets are reported on both MC Doc and MC Psg. 

The statistics of these datasets can be found in Table~\ref{tab:101}. GFLOPs is also attached as a measure of compute.

\paragraph{Implementation}

The finetuning (i.e., the second step of achieving elastic optimization) of \textsc{ElasticLM}, \textsc{ElasticDenser}, and \textsc{ElasticRanker} is mostly the same except that they adopt distinguished task-specific objectives. The finetuning procedure is carried out on one Nvidia A100 GPU. The optimizer is again AdamW. 

For \textsc{ElasticLM} on GLUE, the maximum sequence length is limited 64 for SST-2 but 128 for other tasks. The batch size is searched within a grid \{16,32\}. The learning rate is searched with a grid \{1e-5,2e-5,3e-5\} and the weight decay is set to 1e-2. The number of training epochs is set to 10 with the proportion of warmup epochs being 1e-1. The maximum gradient norm is limited to 5.

For \textsc{ElasticDenser} on NQ and Trivia, the maximum sequence lengths for question and passage are respectively limited to 32 and 156. The batch size is set to 128. The learning rate is set to 1e-5 and the weight decay is set to 0. The number of training epochs is set to 40 with the proportion of warmup epochs being 1e-1. The number of hard negative passages is 1.

For \textsc{ElasticDenser} on MC Psg, the maximum sequence lengths for query and passage are respectively set to 16 and 128. The batch size, learning rate, and weight decay are set to 8, 5e-6 and 0 respectively. The number of training epochs is set to 3 with the proportion of warmup epochs being 1e-1. The number of hard negative passages is 7.

For \textsc{ElasticRanker} on MC Doc, the maximum sequence lengths for concatenated query and document are set to 512. The batch size, learning rate, and weight decay are set to 4, 1e-5 and 0.01 respectively. The number of training epochs is set to 2 with the proportion of warmup epochs being 1e-1. The number of localized negative passages is 7. On this document ranking dataset, we plug \textsc{ElasticRanker} to a retriever named HDCT~\citep{DaiC20} which is once upon a time the state of the art for augmenting document search indices with term weights re-estimated with BERT. For \textsc{ElasticRanker} plugged to \textsc{ElasticDenser} on MC Psg, the settings are not much different except that the maximum sequence length is reduced to 256 regarding that MC Psg has shorter sequences even though the query and the document are concatenated at the input end.

The gathered hands-on details can also be found in Table~\ref{tab:102}

\begin{table}[b]
    \centering
    \caption{The results of \textsc{ElasticLM}. The best results are \underline{underlined}. The comparisons to EarlyExitBERT could be unfair since it uses much more pretraining data. The results with marker \textsuperscript{\S} are significantly better than those of MiniLM in t-test with p-value$<$0.05.}
    \begin{adjustbox}{width=\textwidth,center}
    \begin{tabular}{lll|ccccccc|c}
    \toprule
        \textbf{Method} & \multicolumn{2}{l|}{\textbf{GFLOPs}} & \makecell[c]{\textbf{SST-2}\\\textbf{Acc}} & \makecell[c]{\textbf{MRPC}\\\textbf{F1}} & \makecell[c]{\textbf{STS-B}\\\textbf{SpCorr}} & \makecell[c]{\textbf{QQP}\\\textbf{F1}} & \makecell[c]{\textbf{MNLI-m/mm}\\\textbf{Acc}} & \makecell[c]{\textbf{QNLI}\\\textbf{Acc}} & \makecell[c]{\textbf{RTE}\\\textbf{Acc}} & \makecell[c]{\textbf{GLUE}\\\textbf{Score}} \\
    \midrule
        BERT\textsubscript{\sf base} & 10.9 & \rotatebox[origin=c]{90}{1$\times$} & 93.8 & 91.5 & 87.1 & 88.4 & 84.9/84.9 & 91.9 & 71.5 & 86.7 \\
    \midrule
        KD\textsubscript{\sf 50\%} & 5.44 &  & 92.6 & 89.6 & 87.0 & 88.3 & 83.6/\underline{84.0} & 90.4 & 70.4 & 85.7 \\
        DynaBERT\textsubscript{\sf 50\%} & 5.44 &  & 91.6 & 89.3 & 88.1 & 87.9 & 83.4/83.4 & 90.7 & 65.7 & 85.0 \\
        MoEBERT\textsubscript{\sf 50\%} & 5.44 &  & 92.7 & 90.9 & 87.7 & 88.3 & 83.8/83.9 & 90.6 & 69.7 & 85.9 \\
        TinyBERT\textsubscript{\sf 6L;768H} & 5.44 &  & 92.0 & 90.1 & 89.2 & 87.6 & 82.9/82.9 & 90.5 & \underline{71.1} & 85.8 \\
        MiniLM\textsubscript{\sf 6L;768H} & 5.44 &  & 92.7 & \underline{91.7} & 89.1 & 87.6 & 83.4/83.4 & 90.5 & 70.4 & 86.1 \\
        EarlyExitBERT\textsubscript{\sf 6L} & 5.44 &  & 92.7 & 91.3 & \underline{90.2} & \underline{88.4} & \underline{84.2}/\underline{84.0} & 90.6 & \underline{71.1} & \underline{86.6} \\
        \rowcolor{cyan!20} \textsc{ElasticLM}\textsubscript{\sf 50\%} & 5.44 &  & 93.0 & 91.5 & 89.2 & 88.1 & 83.5/83.5 & 90.7 & 69.3 & 86.1 \\
        \rowcolor{cyan!20} \quad w/o elasticity & 5.44 & \multirow{-8}{*}{\rotatebox[origin=c]{90}{2$\times$}} & \underline{93.1} & 91.3 & 89.0 & 88.1 & 83.4/83.8 & \underline{90.9} & 69.7 & 86.2 \\
    \midrule
        KD\textsubscript{\sf 30\%} & 3.26 &  & 92.0 & 88.9 & 86.8 & 87.8 & 82.2/82.7 & 89.8 & 68.2 & 84.8 \\
        DynaBERT\textsubscript{\sf 30\%} & 3.26 &  & 90.3 & 87.4 & 87.2 & 86.6 & 81.5/81.8 & 89.1 & 66.1 & 83.7 \\
        MoEBERT\textsubscript{\sf 30\%} & 3.26 &  & 91.6 & 90.6 & 86.3 & 87.8 & 82.8/82.9 & 89.9 & 67.9 & 85.0 \\
        TinyBERT\textsubscript{\sf 4L;768H} & 3.59 &  & 91.6 & 89.7 & 88.4 & 87.1 & 81.4/81.5 & 89.4 & \underline{70.0} & 84.9 \\
        MiniLM\textsubscript{\sf 4L;768H} & 3.59 &  & 91.3 & 90.2 & 88.6 & 87.1 & 81.7/81.9 & 89.5 & 67.9 & 84.8 \\
        EarlyExitBERT\textsubscript{\sf 4L} & 3.59 &  & 92.2 & 90.7 & \underline{89.6} & \underline{88.2} & 82.3/82.6 & 89.6 & 69.6 & 85.6 \\
        \rowcolor{cyan!20} \textsc{ElasticLM}\textsubscript{\sf 30\%} & 3.26 &  & 92.7\textsuperscript{\S} & 90.9\textsuperscript{\S} & 88.7\textsuperscript{\S} & 87.5\textsuperscript{\S} & 82.6\textsuperscript{\S}/82.9\textsuperscript{\S} & 90.3\textsuperscript{\S} & \underline{70.0}\textsuperscript{\S} & 85.7\textsuperscript{\S} \\
        \rowcolor{cyan!20} \quad w/o elasticity & 3.26 & \multirow{-8}{*}{\rotatebox[origin=c]{90}{$\sim$3$\times$}} & \underline{92.9} & \underline{91.4} & 89.1 & 87.6 & \underline{82.9}/\underline{83.3} & \underline{90.9} & 69.7 & \underline{86.0} \\
    \midrule
        KD\textsubscript{\sf 15\%} & 1.63 &  & 89.9 & 88.6 & 85.1 & 86.2 & 79.8/80.2 & 85.6 & 63.9 & 82.4 \\
        DynaBERT\textsubscript{\sf 15\%} & 1.63 &  & 89.1 & 85.1 & 84.7 & 84.3 & 78.3/79.0 & 86.6 & 61.4 & 81.1 \\
        MoEBERT\textsubscript{\sf 15\%} & 1.63 &  & 89.6 & 88.4 & 85.1 & \underline{86.8} & \underline{80.4}/\underline{80.5} & 86.6 & 65.0 & 82.8 \\
        TinyBERT\textsubscript{\sf 4L;384H} & 0.91 &  & 89.0 & 88.7 & 86.9 & 86.2 & 79.1/79.6 & 88.2 & \underline{67.5} & 83.1 \\
        MiniLM\textsubscript{\sf 4L;384H} & 0.91 &  & 90.0 & 88.6 & 87.2 & 86.1 & 80.0/80.3 & 87.9 & 67.2 & 83.4 \\
        EarlyExitBERT\textsubscript{\sf 2L} & 1.81 &  & 91.1 & 84.6 & 86.6 & 85.8 & 76.8/76.8 & 84.5 & 63.2 & 81.2 \\
        \rowcolor{cyan!20} \textsc{ElasticLM}\textsubscript{\sf 10\%} & 1.09 &  & 90.5\textsuperscript{\S} & 88.8\textsuperscript{\S} & \underline{88.2}\textsuperscript{\S} & 86.4\textsuperscript{\S} & 80.3\textsuperscript{\S}/80.0 & \underline{88.5}\textsuperscript{\S} & 67.2 & 83.7\textsuperscript{\S} \\
        \rowcolor{cyan!20} \quad w/o elasticity & 1.09 & \multirow{-8}{*}{\rotatebox[origin=c]{90}{6$\sim$12$\times$}} & \underline{92.0} & \underline{90.1} & 87.9 & 86.6 & 80.0/80.3 & 88.0 & 67.2 & \underline{84.0} \\
    \bottomrule
    \end{tabular}
    \end{adjustbox}
    \label{tab:2}
\end{table}

\paragraph{Baselines}

Our baselines range from task-specific to task-agnostic and from once-for-one to once-for-all ones, as detailed below:
\begin{itemize}
    \item KD~\citep{HintonVD15} is a task-specific and once-for-one method that distils logits. This method is only compared to on GLUE.
    \item MoEBERT~\citep{ZuoZLHZC22} is a task-specific and once-for-one method that advances the state-of-the-art performance. This method is only compared to on GLUE.
    \item DynaBERT~\citep{HouHSJCL20} is a task-specific and once-for-all method that distils logits and hidden states. This method is only compared to on GLUE.
    \item TinyBERT~\citep{JiaoYSJCL0L20} is a task-agnostic and once-for-one method that distils hidden states and attention scores. We exclude its task-specific distillation stage for fair comparisons.
    \item MiniLM~\citep{WangBHDW21} is a task-agnostic and once-for-one methods that distils attention distributions.
    \item EarlyExitBERT~\citep{LiuSHWWZJCHQ22} is a task-agnostic and once-for-all method with online scalability yet without online controllability. Since the evaluation is offline, we enforce it to defined compute for comparison.
\end{itemize}

We do not include task-specific baselines for evaluation in dense retrieval, for they are not designed for dense retrieval and can not be directly applied to dense retrieval. Innately, \textsc{ElasticLM} and \textsc{ElasticDenser} separately correspond to their degenerated static counterparts \textsc{ElasticLM} w/o elasticity and \textsc{ElasticDenser} w/o elasticity as ablated baselines. Here, \textsc{ElasticLM} w/o elasticity only differs from MiniLM in that \textsc{MiniLM} uses a random structure rather than a pruned one.

Since reranking is essentially associated with two stages, we here name a few comparable baselines specially for \textsc{ElasticRanker} as below:
\begin{itemize}
    \item BM25+LM-Ranker Distil~\citep{GaoDC20} is a two-stage and once-for-one method that distils logits both on task-agnostic and task-specific data.
    \item BERT+SimTinyBERT~\citep{ChenHHSS21} is also a two-stage and once-for-one method that unifies the distillation on task-agnostic and task-specific data.
\end{itemize}

For all these baselines, structures are denoted either with \textsubscript{\sf *L,*H} for preserved number of layers and hidden dimensions or with \textsubscript{\sf *\%} for preserved portion of parameters in pruning.

\begin{table}[t]
    \centering
    \caption{The results of \textsc{ElasticDenser}. The best results are \underline{underlined}. The comparisons to EarlyExitBERT could be unfair since it uses much more pretraining data. The results with marker \textsuperscript{\S} are significantly better than those of \textsc{ElasticLM} in t-test with p-value$<$0.05.}
    \begin{adjustbox}{width=0.97\textwidth,center}
    \begin{tabular}{lll|ccc|ccc|c}
    \toprule
        \textbf{Method} & \multicolumn{2}{l|}{\textbf{GFLOPs}} & \makecell[c]{\\\textbf{R@5}} & \makecell[c]{\textbf{NQ}\\\textbf{R@20}} & \makecell[c]{\\\textbf{R@100}} & \makecell[c]{\\\textbf{R@5}} & \makecell[c]{\textbf{Trivia}\\\textbf{R@20}} & \makecell[c]{\\\textbf{R@100}} & \makecell[c]{\textbf{MC Psg}\\\textbf{MRR@10}} \\
    \midrule
        BERT\textsubscript{\sf base} & 10.9 &  & 68.6 & 80.3 & 87.6 & 73.0 & 81.4 & 86.4 & 32.0 \\
        Condenser\textsubscript{\sf base} & 10.9 & \multirow{-2}{*}{\rotatebox[origin=c]{90}{1$\times$}} & 70.7 & 81.9 & 88.1 & 75.5 & 82.4 & 86.9 & 34.0 \\
    \midrule
        TinyBERT\textsubscript{\sf 6L;768H} & 5.44 &  & 65.3 & 78.8 & 86.2 & 69.9 & 79.3 & 85.3 & 30.4 \\
        MiniLM\textsubscript{\sf 6L;768H} & 5.44 &  & 66.3 & 79.9 & 87.2 & 72.4 & 80.7 & \underline{86.2} & 31.5 \\
        EarlyExitBERT\textsubscript{\sf 6L} & 5.44 &  & 66.9 & 80.1 & \underline{87.9} & 71.5 & 80.4 & 85.9 & \underline{32.6} \\
        \textsc{ElasticLM}\textsubscript{\sf 50\%} & 5.44 &  & 66.7 & 78.9 & 86.5 & 69.8 & 79.2 & 85.3 & 31.7 \\
        \quad w/o elasticity & 5.44 &  & 66.8 & 79.3 & 87.2 & 70.8 & 80.0 & 85.6 & 31.0 \\
        \rowcolor{cyan!20} \textsc{ElasticDenser}\textsubscript{\sf 50\%} & 5.44 &  & 67.2\textsuperscript{\S} & 79.6\textsuperscript{\S} & 87.0\textsuperscript{\S} & 71.1\textsuperscript{\S} & 80.0\textsuperscript{\S} & 85.6\textsuperscript{\S} & 32.4\textsuperscript{\S} \\
        \rowcolor{cyan!20} \quad w/o elasticity & 5.44 & \multirow{-7}{*}{\rotatebox[origin=c]{90}{2$\times$}} & \underline{68.5} & \underline{80.5} & 87.4 & \underline{73.2} & \underline{81.1} & 86.1 & \underline{32.6} \\
    \midrule
        TinyBERT\textsubscript{\sf 4L;768H} & 3.59 &  & 63.3 & 77.6 & 85.5 & 67.9 & 77.9 & 84.6 & 29.0 \\
        MiniLM\textsubscript{\sf 4L;768H} & 3.59 &  & 66.6 & 78.9 & 86.7 & \underline{71.4} & \underline{80.3} & \underline{86.0} & 30.7 \\
        EarlyExitBERT\textsubscript{\sf 4L} & 3.59 &  & 65.3 & 78.7 & 86.8 & 70.6 & 79.8 & 85.4 & 31.4 \\
        \textsc{ElasticLM}\textsubscript{\sf 30\%} & 3.26 &  & 66.3 & 78.9 & 86.5 & 69.4 & 78.6 & 85.0 & 31.6 \\
        \quad w/o elasticity & 3.26 &  & 66.4 & 79.3 & 86.8 & 69.8 & 79.3 & 85.3 & 31.0 \\
        \rowcolor{cyan!20} \textsc{ElasticDenser}\textsubscript{\sf 30\%} & 3.26 &  & 66.6\textsuperscript{\S} & 79.2\textsuperscript{\S} & 87.0\textsuperscript{\S} & 70.3\textsuperscript{\S} & 79.6\textsuperscript{\S} & 85.4\textsuperscript{\S} & \underline{32.2}\textsuperscript{\S} \\
        \rowcolor{cyan!20} \quad w/o elasticity & 3.26 & \multirow{-7}{*}{\rotatebox[origin=c]{90}{$\sim$3$\times$}} & \underline{67.6} & \underline{79.8} & \underline{87.2} & 71.0 & 80.2 & 85.5 & 31.6 \\
    \midrule
        TinyBERT\textsubscript{\sf 4L;384H} & 0.91 &  & 54.3 & 70.6 & 81.4 & 55.8 & 69.8 & 79.5 & 22.6 \\
        MiniLM\textsubscript{\sf 4L;384H} & 0.91 &  & 61.5 & 75.6 & 84.7 & 65.3 & 76.0 & 83.4 & 27.4 \\
        EarlyExitBERT\textsubscript{\sf 2L} & 1.81 &  & 59.4 & 75.4 & 84.6 & 64.6 & 75.9 & 83.4 & 27.5 \\
        \textsc{ElasticLM}\textsubscript{\sf 10\%} & 1.09 &  & 61.4 & 75.9 & 84.7 & 64.6 & 75.7 & 83.4 & 29.6 \\
        \quad w/o elasticity & 1.09 &  & 61.0 & 75.8 & 84.9 & 64.8 & 75.7 & 83.2 & 28.6 \\
        \rowcolor{cyan!20} \textsc{ElasticDenser}\textsubscript{\sf 10\%} & 1.09 &  & \underline{64.0}\textsuperscript{\S} & \underline{77.4}\textsuperscript{\S} & \underline{86.1}\textsuperscript{\S} & \underline{67.1}\textsuperscript{\S} & \underline{77.6}\textsuperscript{\S} & \underline{84.0}\textsuperscript{\S} & \underline{30.7}\textsuperscript{\S} \\
        \rowcolor{cyan!20} \quad w/o elasticity & 1.09 & \multirow{-7}{*}{\rotatebox[origin=c]{90}{6$\sim$12$\times$}} & 62.1 & 76.6 & 85.3 & 66.0 & 77.0 & 83.7 & 28.8 \\
    \bottomrule
    \end{tabular}
    \end{adjustbox}
    \label{tab:3}
\end{table}

\paragraph{Results}
\label{sec:1}

From results of \textsc{ElasticLM} as shown in Table~\ref{tab:2}, we can observe that both \textsc{ElasticLM} and its static counterpart \textsc{ElasticLM} w/o elasticity outperform almost all FLOPs-matched baselines. For example, \textsc{ElasticLM}\textsubscript{\sf 30\%} w/o elasticity shows a 1.2 absolute performance improvement over MiniLM\textsubscript{\sf 4L;768H} in terms of GLUE Score. This hints that pruning, as a key to deriving the elastic structure, can enhance the performance of task-agnostic distillation. It is noteworthy that EarlyExitBERT is not directly comparable since it uses much more pretraining data. Plus, we can see that \textsc{ElasticLM} yields competitive performance in comparison with \textsc{ElasticLM} w/o elasticity. For example, \textsc{ElasticLM}\textsubscript{\sf 50\%} yields an 86.1 against 86.2 of that w/o elasticity in terms of GLUE Score. This indicates that the elastic optimization would not degrade the performance that much. 

From results of \textsc{ElasticDenser} as displayed in Table~\ref{tab:3}, we can further validate the findings discovered on GLUE. Moreover, we find that the elastic optimization can bring performance gains for both \textsc{ElasticLM} over \textsc{ElasticLM} w/o elasticity and \textsc{ElasticDenser} over \textsc{ElasticDenser} w/o elasticity. For example, \textsc{ElasticDenser}\textsubscript{\sf 10\%} outperforms that w/o elasticity by a 1.9 absolute margin. Without much surprise, we also reveal that \textsc{ElasticDenser} is a more promising choice than \textsc{ElasticLM} in the context of dense retrieval. Here, \textsc{ElasticLM} is finetuned towards dense retrieval using tactics similar to those used by \textsc{ElasticDenser}.

From results of \textsc{ElasticRanker} in Table~\ref{tab:103}, we can safely say that \textsc{ELasticRanker} is more promising than that w/o elasticity. For example, \textsc{ElasticRanker}\textsubscript{\sf 10\%} outperforms \textsc{ElasticRanker}\textsubscript{\sf 10\%} w/o elasticity and baselines. We have to permit the baselines are sort of weak to be compared to, but we have already tried our best to dig out the most comparable results.

The results of full possible parameter-preserving levels, including various combinations of \textsc{ElasticDenser} and \textsc{ElasticRanker}, are deferred to Section~\ref{sec:2}.

Overall, these results show that \textsc{ElasticLM}, \textsc{ElasticDenser}, and \textsc{ELasticRanker} perform correctly with compute elasticity and competitively with the static baselines.

\begin{table}[t]
    \centering
    \caption{The results of \textsc{ElasticRanker}. The best results are \underline{underlined}. The results with marker \textsuperscript{\S} are significantly better than those of BERT+SimTinyBERT or BM25+LM-Ranker Distil in t-test with p-value$<$0.05.}
    \begin{adjustbox}{width=0.99\textwidth,center}
    \begin{tabular}{lll|cc|cc}
    \toprule
        \multirow{2}{*}{\textbf{Method}} & \multicolumn{2}{l|}{\multirow{2}{*}{\textbf{GFLOPs}}} & \multicolumn{2}{c|}{\textbf{MC Doc}} & \multicolumn{2}{c}{\textbf{MC Psg}} \\
         &  &  & \textbf{MRR@10} & \textbf{MRR@100} & \textbf{MRR@10} & \textbf{MRR@100} \\
    \midrule
        HDCT\textsubscript{\sf base}$+$BERT\textsubscript{\sf base} & 10.9 & \multirow{2}{*}{\rotatebox[origin=c]{90}{1$\times$}} & 42.1 & 42.8 & - & - \\
        HDCT\textsubscript{\sf base}$+$PROP\textsubscript{\sf base} & 10.9 &  & 42.8 & 43.5 & - & - \\
    \midrule
        BERT\textsubscript{\sf base}+SimTinyBERT\textsubscript{\sf 6L;768H} & 5.44 & & 38.5 & 39.1 & - & - \\
        \rowcolor{cyan!20} HDCT\textsubscript{\sf base}$+$\textsc{ElasticRanker}\textsubscript{\sf 50\%} & 5.44 &  & \underline{42.9}\textsuperscript{\S} & \underline{43.6}\textsuperscript{\S} & - & - \\
        \rowcolor{cyan!20} \quad w/o elasticity & 5.44 & \multirow{-3}{*}{\rotatebox[origin=c]{90}{2$\times$}} & 42.5 & 43.2 & - & - \\
    \midrule
        BM25+LM-Ranker Distil\textsubscript{\sf 6L;768H} & 5.44 &  & - & - & 36.0 & - \\
        \rowcolor{cyan!20} \textsc{ElasticDenser}\textsubscript{\sf 50\%}+\textsc{ElasticRanker}\textsubscript{\sf 50\%} & 5.44 & \multirow{-2}{*}{\rotatebox[origin=c]{90}{2$\times$}} & - & - & \underline{39.6}\textsuperscript{\S} & \underline{40.5}\textsuperscript{\S} \\
    \midrule
        BERT\textsubscript{\sf base}+SimTinyBERT\textsubscript{\sf 3L;384H} & 0.68 &  & 36.1 & 36.8 & - & - \\
        \rowcolor{cyan!20} HDCT\textsubscript{\sf base}$+$\textsc{ElasticRanker}\textsubscript{\sf 10\%} & 1.09 &  & \underline{42.0}\textsuperscript{\S} & \underline{42.7}\textsuperscript{\S} & - & - \\
        \rowcolor{cyan!20} \quad w/o elasticity & 1.09 & \multirow{-3}{*}{\rotatebox[origin=c]{90}{10$\sim$16$\times$}} & 41.7 & 42.4 & - & - \\
    \midrule
        BM25+LM-Ranker Distil\textsubscript{\sf 4L;768H} & 3.59 &  & - & - & 35.0 & - \\
        \rowcolor{cyan!20} \textsc{ElasticDenser}\textsubscript{\sf 10\%}$+$\textsc{ElasticRanker}\textsubscript{\sf 10\%} & 1.09 & \multirow{-2}{*}{\rotatebox[origin=c]{90}{3$\sim$9$\times$}} & - & - & \underline{38.7}\textsuperscript{\S} & \underline{39.6}\textsuperscript{\S} \\
    \bottomrule
    \end{tabular}
    \end{adjustbox}
    \label{tab:103}
\end{table}

\subsection{Online Simulation}

\paragraph{Data \& Metrics}

To mimic real-world online requests, we randomly sample a number of queries from MC Psg as pseudo queries in our simulation. During the online simulation, we mainly keep an eye on the latency and the performance. 

\begin{figure}[t]
    \centering
    \includegraphics[width=0.82\textwidth]{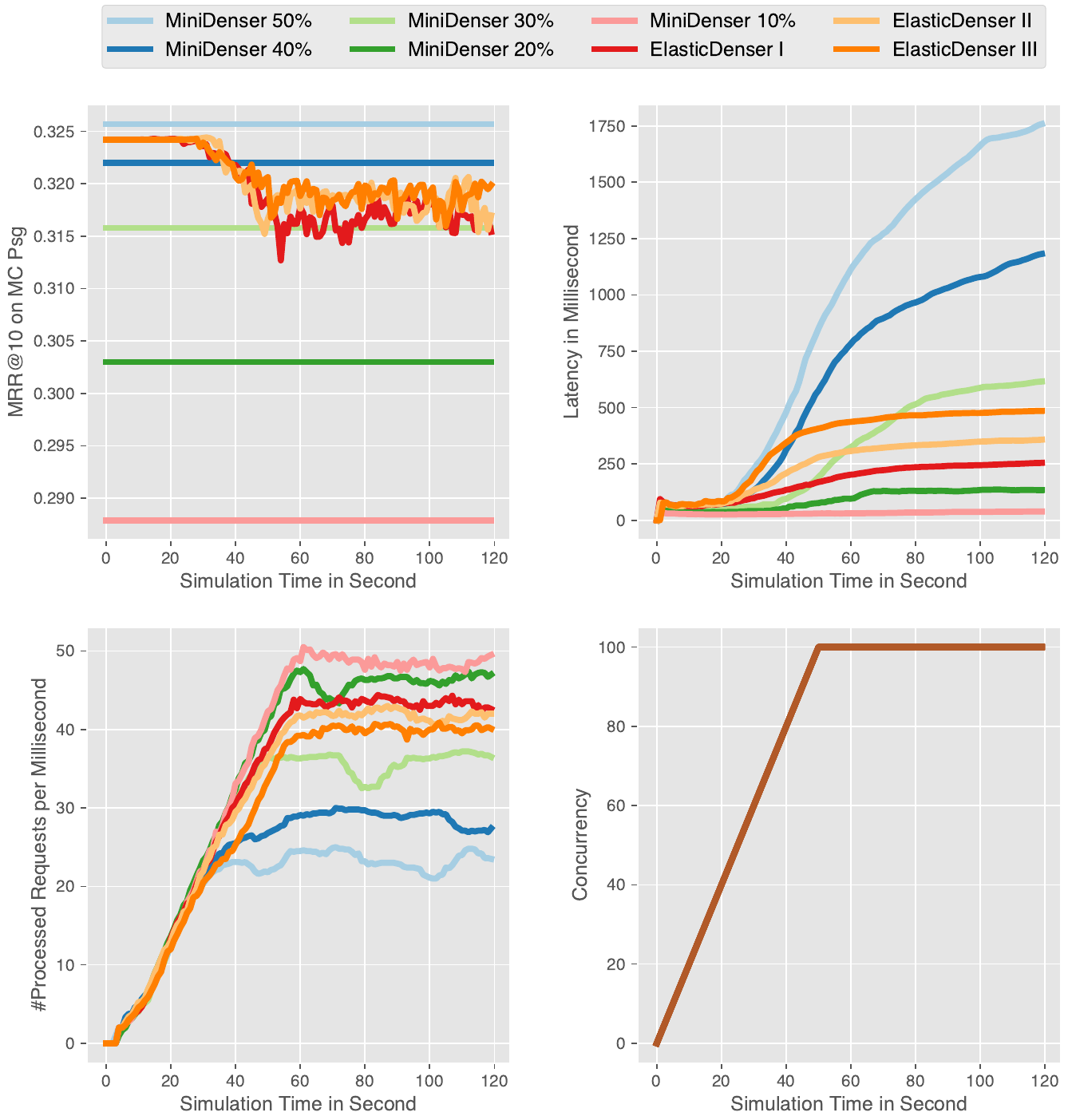}
    \caption{The results of online simulation. Best viewed in color.}
    \label{fig:7}
\end{figure}

\paragraph{Implementation}

The simulation is conducted on an Intel Core i7 CPU. The duration of the simulation is limited to 120 seconds, where the concurrency is linearly increased to 100 in the first 50 seconds (i.e., increasing 2 per second) and kept unchanged in the rest of time. In doing so, we can establish a simple rush case as illustrated in the bottom right of Figure~\ref{fig:7}. Each routine is driven by a load testing toolkit Locust,\footnote{\url{https://github.com/locustio/locust}} and the routine could continuously send requests with a pause of one to three seconds inserted between two sequential requests.

Following the design regime of a model serving toolkit TorchServe,\footnote{\url{https://github.com/pytorch/serve}} \textsc{ElasticDenser} is served with a request queue, where new requests will be appended to the queue tail and \textsc{ElasticDenser} will each time process one request popped from the queue head. 

For simplicity, we waive the processing time consumed by passage searching and mainly focus on the processing time consumed by query encoding. We admit this is a far unpractical way to conduct simulation for \textsc{ElasticDenser}, however, we argue that the passage searching procedure would not be accelerated by our method since \textsc{ElasticLM} is concerned with model size instead of embedding size. And this simplification allows us to show the significance of the core contribution of \textsc{ElasticLM}.

\paragraph{Baselines}

We compare \textsc{ELasticDenser} with its static counterparts, i.e., \textsc{ELasticDenser} w/o elasticity with preserving levels \{50,40,30,20,10\}\%. We also examine the effect of latency constraint (i.e., $T$) by varying it within \{250,375,500\} milliseconds. Accordingly, we name the corresponding \textsc{ELasticDenser}s as \textsc{ELasticDenser} \{I,II,III\}.

\paragraph{Results}

As shown in Figure~\ref{fig:7}, compared to static \textsc{MiniDenser} (i.e., \textsc{ElasticDenser} w/o elasticity mentioned above for brevity), \textsc{ElasticDenser} demonstrates a capability of elastically adjusting itself to keep a good latency-performance tradeoff given the latency constraint. In contrast, large \textsc{MiniDenser} would explode to high latency beyond the constraint, while small \textsc{MiniDenser} would wander around low performance below the expectation. These results show that the inner working of \textsc{ElasticLM} enables appealing tradeoffs.

\subsection{Analyses}
\label{sec:2}

\paragraph{Preserving Levels}

To obtain a more complete understanding of how parameter-preserving level could impact the performance, we further plot the performance variation along the reduction of the preserving level.

\begin{figure}[h]
    \centering
    \includegraphics[width=\textwidth]{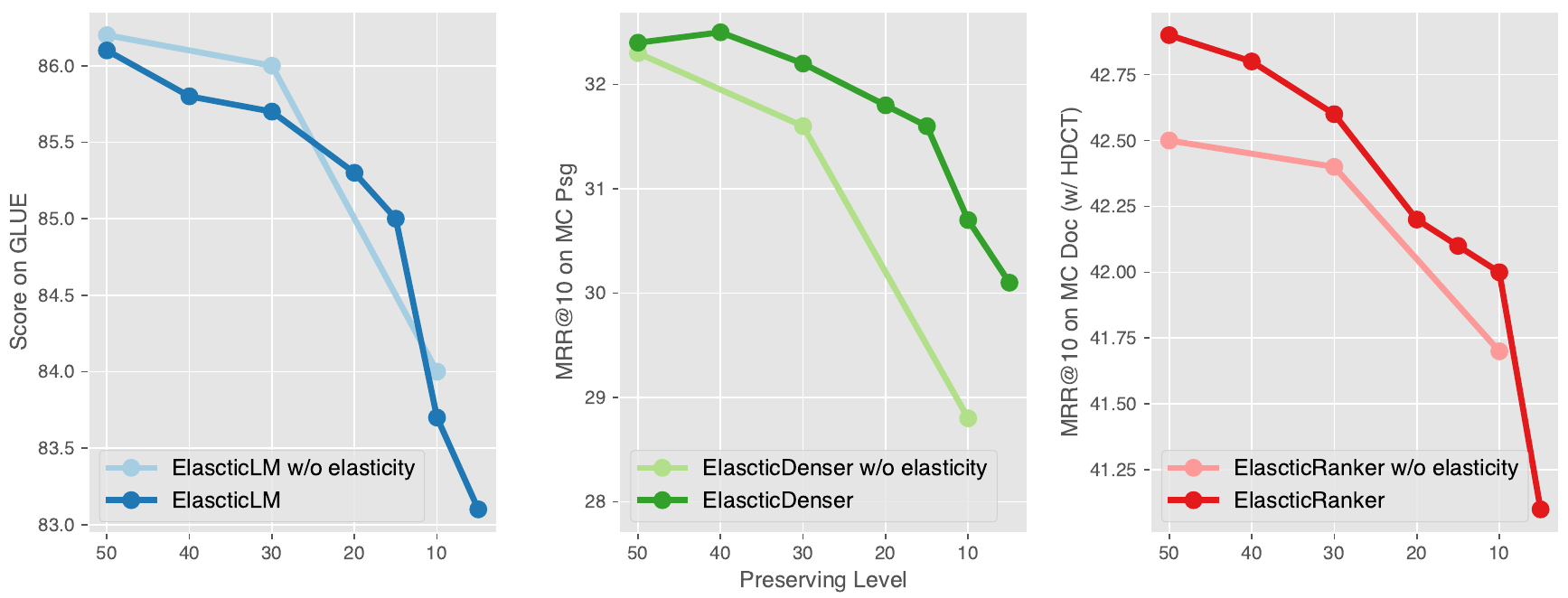}
    \caption{The performance variations of \textsc{ElasticLM}, \textsc{ElasticDenser}, and \textsc{ElasticRanker} along the reduction of preserving levels.}
    \label{fig:103}
\end{figure}

\begin{figure}[t]
    \centering
    \includegraphics[width=0.52\textwidth]{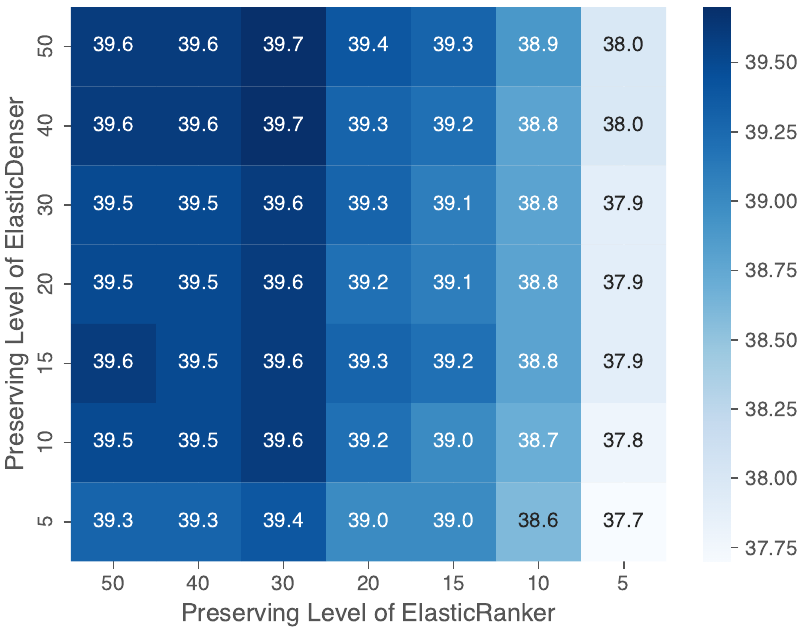}
    \caption{The overall performance variations of \textsc{ElasticRanker} plugged to \textsc{ElasticDenser} as a pipeline along the reduction of preserving levels.}
    \label{fig:104}
\end{figure}

From Figure~\ref{fig:103}, we get similar but more complete trends as those in Section~\ref{sec:1} that \textsc{ElasticLM}, \textsc{ElasticDenser}, and \textsc{ElasticRanker} could keep competitive with and sometimes even outperforms their static counterparts. For example, \textsc{ElasticDenser} is entirely superior to \textsc{ElasticDenser} w/o elasticity. At the same time, when preserving level is reduced, the performance would not degrade that much, verifying the elastic optimization performs correctly for the second time.

From Figure~\ref{fig:104}, when \textsc{ElasticRanker} is plugged to \textsc{ElasticDenser}, it seems that variations in preserving level of \textsc{ElasticDenser} (variations within each column) would has little impact on the overall performance of the whole pipeline . And even variations in preserving level of \textsc{ElasticRanker} (variations within each row) can give notable performance impact, the performance deline still lies in an acceptable range, implying that the elastic pipeline is of interest.

\paragraph{Distillation Variants}

\begin{figure}[t]
    \centering
    \includegraphics[width=0.7\textwidth]{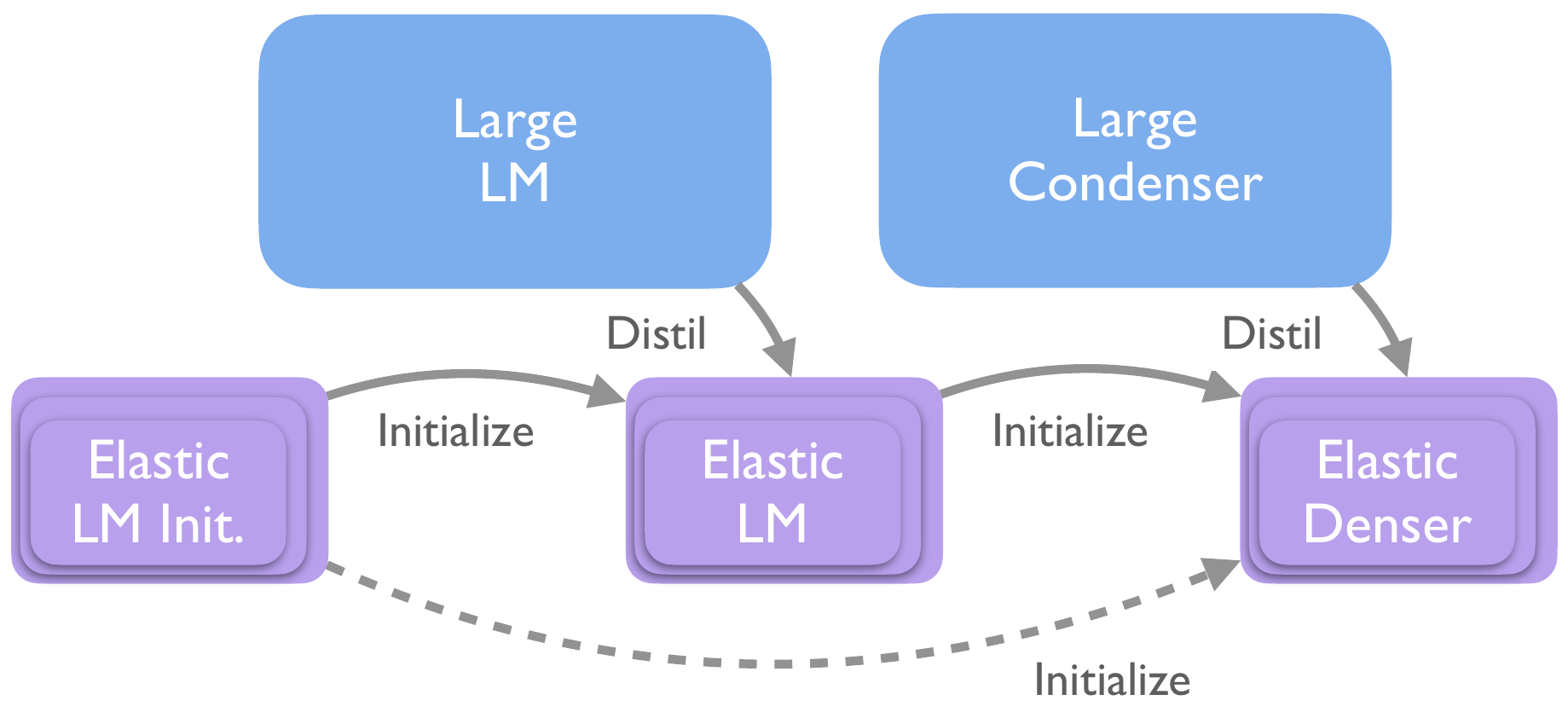}
    \caption{The illustration of \textsc{ElasticDenser} w/o and w/ cascade. \textsc{ElasticDenser} w/ cascade requires initialization from trained \textsc{ElasticLM}.}
    \label{fig:101}
\end{figure}

\begin{figure}[t]
    \centering
    \includegraphics[width=0.99\textwidth]{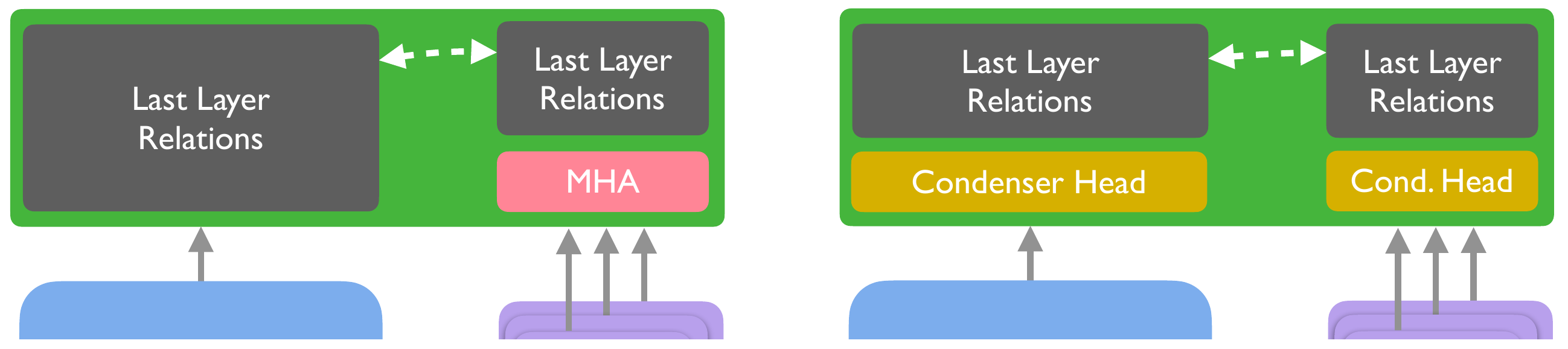}
    \caption{The illustration of \textsc{ElasticDenser} w/o (left) and w/ (right) head. \textsc{ElasticDenser} w/ head requires reuse of trained head from Condenser.}
    \label{fig:102}
\end{figure}

We study two possible variants of \textsc{ElasticDenser} during task-agnostic distillation. The first (\textsc{ElasticDenser} w/ cascade) is to initialize the elastic structure with that from the optimized \textsc{ElasticLM} rather than that directly from pruned BERT as in Figure~\ref{fig:101}. The second (\textsc{ElasticDenser} w/ head) is to align the relations of the last MHA block from the Condenser head instead of directly the Condenser backbone as in Figure~\ref{fig:102}.

\begin{table}[h]
    \centering
    \caption{The results of distillation variants.}
    \begin{adjustbox}{width=0.6\textwidth,center}
    \begin{tabular}{l|ccc|c}
    \toprule
        \textbf{Method} & \makecell[c]{\\\textbf{R@5}} & \makecell[c]{\textbf{NQ}\\\textbf{R@20}} & \makecell[c]{\\\textbf{R@100}} & \makecell[c]{\textbf{MARCO}\\\textbf{MRR@10}} \\
    \midrule
        \textsc{ElasticDenser}\textsubscript{\sf 30\%} & 66.6 & 79.2 & 87.0 & 32.2 \\
        \quad w/ cascade & 66.7 & 79.5 & 87.0 & 31.9 \\
        \quad w/ head & 63.6 & 78.2 & 85.8 & 30.7 \\
    \midrule
        \textsc{ElasticDenser}\textsubscript{\sf 10\%} & 64.0 & 77.4 & 86.1 & 30.7 \\
        \quad w/ cascade & 61.6 & 76.3 & 85.6 & 29.8 \\
        \quad w/ head & 58.7 & 73.9 & 83.7 & 28.2 \\
    \bottomrule
    \end{tabular}
    \end{adjustbox}
    \label{tab:4}
\end{table}

From the results in Table~\ref{tab:4}, we notice that \textsc{ElasticDenser} w/ cascade can achieve similar performance to \textsc{ElasticDenser} at a slightly large scale but lag behind \textsc{ElasticDenser} at a slightly small scale. We also mark that \textsc{ElasticDenser} w/ head is not an ideal design choice as its performance 
deteriorates significantly from that of \textsc{ElasticDenser}.

\paragraph{Large Teachers}

We inspect the effect of distillation from large teachers such as  Condenser\textsubscript{\sf large}. Additionally, we explore how the idea of progressive distillation~\citep{Lin22} can help the distillation from the large teacher (e.g., first distilling from Condenser\textsubscript{\sf base} then distilling from Condenser\textsubscript{\sf large}).

\begin{table}[h]
    \centering
    \caption{The results of distillation from large teachers.}
    \begin{adjustbox}{width=0.65\textwidth,center}
    \begin{tabular}{l|ccc|c}
    \toprule
        \textbf{Method} & \makecell[c]{\\\textbf{R@5}} & \makecell[c]{\textbf{NQ}\\\textbf{R@20}} & \makecell[c]{\\\textbf{R@100}} & \makecell[c]{\textbf{MARCO}\\\textbf{MRR@10}} \\
    \midrule
        Condenser\textsubscript{\sf base} & 70.7 & 81.9 & 88.1 & 34.0 \\
        $\Rightarrow$\textsc{ElasticDenser}\textsubscript{\sf 50\%} & 67.2 & 79.6 & 87.0 & 32.4 \\
        Condenser\textsubscript{\sf large} & 73.0 & 82.9 & 88.6 & 35.8 \\
        $\Rightarrow$\textsc{ElasticDenser}\textsubscript{\sf 50\%} & 65.2 & 78.8 & 86.7 & 31.8 \\
        $\Rightarrow\Rightarrow$\textsc{ElasticDenser}\textsubscript{\sf 50\%} & 66.0 & 78.5 & 86.7 & 31.8 \\
    \bottomrule
    \end{tabular}
    \end{adjustbox}
    \label{tab:5}
\end{table}

The results in Table~\ref{tab:5} show that large teachers could invoke worse students. While this is a already known curse~\citep{MirzadehFLLMG20}, it seems more severe for dense retrieval in term of the critical performance drop. Despite that, the results show that progressive distillation can partially lift the performance.

\paragraph{Random versus Pruning}

We investigate whether random initialization could replace the pruning initialization for the elastic structure, where random initialization admits manual setting of structures (e.g., the number of layers and the hidden dimension, etc.).

The results in Table~\ref{tab:6} suggest that pruning outperforms random by large margins, implying the use of pruning for the elastic structure is beneficial.

\begin{table}[h]
    \centering
    \caption{The results of random versus pruning. Random initialization admits manual setting of structures. {\sf 4L;768H} is a compute-equivalent random structure to {\sf 30\%} in the case.}
    \begin{adjustbox}{width=0.62\textwidth,center}
    \begin{tabular}{l|ccc|c}
    \toprule
        \textbf{Method} & \makecell[c]{\\\textbf{R@5}} & \makecell[c]{\textbf{NQ}\\\textbf{R@20}} & \makecell[c]{\\\textbf{R@100}} & \makecell[c]{\textbf{MARCO}\\\textbf{MRR@10}} \\
    \midrule
        \textsc{ElasticLM}\textsubscript{\sf 30\%} & 66.3 & 78.9 & 86.5 & 31.6 \\
        \textsc{ElasticLM}\textsubscript{\sf 4L;768H} & 64.2 & 77.4 & 85.7 & 29.9 \\
    \bottomrule
    \end{tabular}
    \end{adjustbox}
    \label{tab:6}
\end{table}

\section{Conclusions and Future Work}

In this paper, we argue conventional language models owning static latency-performance tradeoffs do not always fit for the scenarios where the number of requests is highly variant (e.g., IR). To tackle the problem, we propose an \textsc{ElasticLM} based on compute elasticity, and design an elastic structure, an elastic optimization, and an elastic schedule to achieve it. \textsc{ElasticLM} is then adapted to dense retrieval and reranking. The experimental results of both offline evaluation and online simulation signal that \textsc{ElasticLM} performs correctly and competitively with expected elastic tradeoffs. 

Our online simulation is a rather naive and simplified one. In the future, we plan to serve \textsc{ElasticDenser} and \textsc{ElasticRanker} as a complete pipeline for IR and test the pipeline under real-world concurrency. 




\bibliographystyle{acm}
\bibliography{sample-base}

\appendix

\end{document}